\documentclass[sigconf]{acmart}

\AtBeginDocument{%
  }

\copyrightyear{2025} 
\acmYear{2025} 
\setcopyright{acmlicensed}\acmConference[CHI '25]{CHI Conference on Human Factors in Computing Systems}{April 26-May 1, 2025}{Yokohama, Japan}
\acmBooktitle{CHI Conference on Human Factors in Computing Systems (CHI '25), April 26-May 1, 2025, Yokohama, Japan}
\acmDOI{10.1145/3706598.3714323}
\acmISBN{979-8-4007-1394-1/25/04}

\usepackage{subcaption}
\usepackage{multirow}
\usepackage[utf8]{inputenc}
\usepackage{natbib}

\begin{document}

\title
[Augmented Journeys: Interactive POIs for In-Car AR]
{Augmented Journeys: Interactive Points of Interest for In-Car Augmented Reality}

\author{Robin Connor Schramm}
\orcid{0000-0002-4775-4219}
\affiliation{
  \institution{Mercedes-Benz Tech Motion GmbH}
  \city{B{\"o}blingen}
  \country{Germany}
}
\affiliation{
  \institution{RheinMain University of Applied Sciences}
  \city{Wiesbaden}
  \country{Germany}
}
\email{robin.schramm@mercedes-benz.com}

\author{Ginevra Fedrizzi}
\orcid{0009-0001-2123-299X}
\affiliation{
  \institution{Mercedes-Benz Tech Motion GmbH}
  \city{B{\"o}blingen}
  \country{Germany}
}
\email{ginevrafedrizzi@gmail.com}

\author{Markus Sasalovici}
\orcid{0000-0001-9883-2398}
\affiliation{
  \institution{Mercedes-Benz Tech Motion GmbH}
  \city{B{\"o}blingen}
  \country{Germany}
}
\affiliation{
  \institution{Ulm University}
  \city{Ulm}
  \country{Germany}
}
\email{markus.sasalovici@mercedes-benz.com}

\author{Jann Philipp Freiwald}
\orcid{0000-0002-1977-5186}
\affiliation{
  \institution{Mercedes-Benz Tech Motion GmbH}
  \city{B{\"o}blingen}
  \country{Germany}
}
\email{jann_philipp.freiwald@mercedes-benz.com}

\author{Ulrich Schwanecke}
\orcid{0000-0002-0093-3922}
\affiliation{
  \institution{RheinMain University of Applied Sciences}
  \city{Wiesbaden}
  \country{Germany}
}
\email{ulrich.schwanecke@hs-rm.de}

\renewcommand{\shortauthors}{Schramm et al.}

\begin{abstract}
    As passengers spend more time in vehicles, the demand for non-driving related tasks (NDRTs) increases. In-car Augmented Reality (AR) has the potential to enhance passenger experiences by enabling interaction with the environment through NDRTs using world-fixed Points of Interest (POIs). However, the effectiveness of existing interaction techniques and visualization methods for in-car AR remains unclear. Based on a survey (N=110) and a pre-study (N=10), we developed an interactive in-car AR system using a video see-through head-mounted display to engage with POIs via eye-gaze and pinch. Users could explore passed and upcoming POIs using three visualization techniques: List, Timeline, and Minimap. We evaluated the system's feasibility in a field study (N=21). Our findings indicate general acceptance of the system, with the List visualization being the preferred method for exploring POIs. Additionally, the study highlights limitations of current AR hardware, particularly the impact of vehicle movement on 3D interaction.
\end{abstract}



\begin{teaserfigure}
  \includegraphics[width=\textwidth]{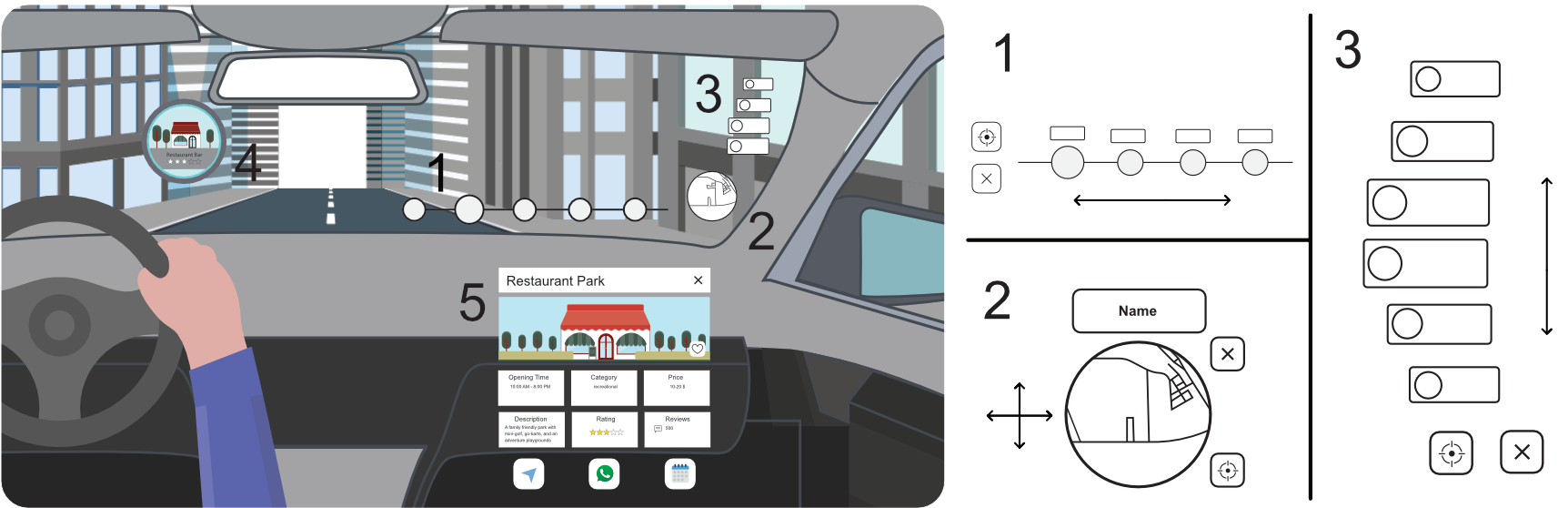}
  \caption{An overview of our system to explore POIs using in-car AR as a passenger. We investigate the interaction with POIs (4) in the environment by displaying additional information to the user (5). We also tested three distinct visualization techniques to explore missed and upcoming POIs: Timeline (1), Minimap (2), and List (3). Each of the three techniques is intended to work in tandem with the world-fixed POIs (4) and the additional POI information (5).}
  \Description{The image depicts an augmented reality (AR) navigation system integrated into a vehicle, providing detailed information about points of interest (POIs) as the driver navigates the road. The system overlays AR elements on the windshield, allowing the driver to view and interact with POIs directly in their line of sight. In the left section of the image, the passengers perspective through the windshield is depicted. The road ahead is augmented with a circular marker, representing an upcoming POI marked with the number four. The system also overlays a timeline numbered one, a minimap numbered two, and a vertical list numbered three in front of the passenger. On the dashboard, there is an additional interface numbered five. This provides further detailed information about the POI,  including details such as the opening hours, the category of the POI (in this case, a restaurant), price range, a description of the location, a rating shown with stars, and a brief review summary. To the right of the image, there are several diagrams that further explain how the AR interface works. The first diagram illustrates a horizontal scrolling interface, depicting a timeline. The second diagram shows a circular minimap which can be scrolled in any direction indicated by arrows. The third diagram shows a vertical list of POIs. Arrows indicate that users can scroll up and down through the list.}
  \label{fig:teaser}
\end{teaserfigure}

\maketitle

\section{Introduction}

Exploring unfamiliar places and accessing relevant information about the surrounding environment are crucial for enhancing passenger engagement and satisfaction. In such situations, individuals often look for intuitive ways to discover nearby Points of Interest (POIs), such as food spots, parks, museums, or gas stations \cite{Psyllidis2022POIs, sun2023conflating}. Providing a simple and effective mechanism to explore the environment and gain easy access to detailed information about these locations can empower users to make informed decisions and deepen their connection with their surroundings. Potential application areas include supporting users in their daily routines and encouraging exploration in unfamiliar cities.

Advances in transportation such as improved public transit and the emergence of autonomous vehicles are expected to shift the role of individuals from drivers to passengers \cite{McGill2022MRPassengerXP, MatsumuraActivePassengering18}.
This transition is anticipated to increase the time individuals spend as passengers, which can be considered as unproductive or wasted time when not used for non-driving related tasks (NDRTs) \cite{gardner2007drives, watts2008moving, wilfinger2011we}.
The potential benefits of NDRTs are vast, offering passengers the ability to partake in activities ranging from work to leisure, thus turning travel time into a productive or enjoyable experience \cite{Mathis2021work, medeiros2022shielding, Togwell2022gaming, Pfleging16NDRNeeds, Riegler19WindshieldArNDRT}. Additionally, passengers often enjoy interacting with their surroundings and appreciating the view through the windows \cite{russell2011passengers, hecht2020ndrts, MatsumuraActivePassengering18,BergerGridStudyInCarPassenger2021}. Recognizing this, one promising approach to NDRTs is to enhance this engagement with the external scenery \cite{MatsumuraActivePassengering18} and to relate it to the current trip \cite{Inbar11TripUX}.
In line with this, Augmented Reality (AR) emerges as a technology that can significantly aid passengers in interacting with their environment, thereby offering a novel way to improve experiences in cars \cite{McGill2022MRPassengerXP}. AR hardware continues to advance both in Head-Up Displays (HUDs) and in Head-Mounted Displays (HMDs), making them more practical for in-car use \cite{riegler2021augmented, Elhattab23AutomotiveAR, Goedicke2022xroom}. Such technologies could enrich the passenger experience by providing access to information about the outside environment \cite{BergerGridStudyInCarPassenger2021}, for instance, through POIs.
Future in-car AR systems could incorporate POIs to present real-time, context-specific information about the car's surroundings overlaid onto the real-world. This seamless integration of digital content with the surrounding context could facilitate the understanding of information at a glance \cite{haeuslschmid2016design} and help passengers to connect more deeply with their environment \cite{BergerGridStudyInCarPassenger2021, Mehrabian74EnvironmentalPsychology, Pfleging16NDRNeeds, Berger21InteractiveCarDoor}.

To allow passengers to interact with their environment, we developed and evaluated an interactive in-car system utilizing an AR HMD in a moving vehicle. Our system allows passengers to explore and interact with world-fixed POIs in their immediate surroundings, as well as explore passed and upcoming POIs. We focus on the exploration of an unknown city-environment, as these environments contain the most POIs. In a preliminary study (N = 10), we evaluated the interaction technique eye-gaze combined with a hardware button in AR settings with world-fixed objects in transit regarding workload and usability. We also conducted a survey (N = 110) on the current behaviours of passengers interacting with location based data in the car without AR. We then adapted our prototype to adress these needs and habits of passengers. To evaluate our system, we conducted a within-subjects user study (N = 21) in a moving vehicle in the field. Here we examined the impact of three seat-fixed spatial user interface (UI) concepts in addition to the previously studied world-fixed POIs, as shown in Figure \ref{fig:teaser}. We evaluated this system regarding workload, usability, user experience and user-preference, together with qualitative data collected through semi-structured interviews. As such, the main contributions of our paper are as follows: 
\begin{itemize}
    \item A comprehensive investigation of the habits and preferences of passengers regarding their current interaction with location-based data in cars through a survey (N=110).
    \item Assessing the feasibility of eye-gaze-based interaction methods for in-car AR through quantitative and qualitative data (N=10 \& N=21).
    \item Evaluation of a system for interacting with POIs and displaying further information using AR through quantitative data and semi-structured interviews (N=21).
\end{itemize}
\section{Related work}
\label{section:related}
Our work explores the visualization and interaction with location based data for in-car AR. As such, our related work consists of the topics of augmented passenger experiences, in-car AR interaction, and visualization of location-based data in vehicles.

\subsection{Augmented Passenger Experiences}
Existing work highlights the potential of AR to enhance passenger experiences, particularly for infotainment and interaction with the external environment. However, significant challenges in this field remain.

There has been extensive research investigating the use of AR to facilitate NDRTs during transit. While NDRTs have been studied extensively, most research has focused on drivers and future scenarios involving automated vehicles \cite{Pfleging16NDRNeeds}. Pfleging et al. \cite{Pfleging16NDRNeeds} explored NDRT related activities drivers wish to engage in during highly or fully automated vehicle operation,  revealing a strong interest in daydreaming, texting, browsing the Internet, using mobile multimedia applications, and watching the environment. Regarding passenger experiences, Berger et al. \cite{BergerGridStudyInCarPassenger2021} identified well-being, physical comfort, and safety as critical influencing factors. Their study also highlighted that passengers value access to in-vehicle systems, the ability to act as a co-driver, and the integration of external technology for connectivity and personalization. Additionally, the importance of the outside environment was emphasized, with participants preferring routes with scenic views to enhance their ride experience and maintain situational awareness.

Some works have evaluated to include AR to improve passenger experiences. Togwell et al. \cite{Togwell2022gaming} explored the potential of integrating AR gaming with the sensing capabilities of autonomous vehicles, enabling games that incorporate real-world elements, such as other cars, into the gameplay. Initial findings suggest that in-car AR gaming enhances passenger experience and immersion by using the vehicle's environment creatively, offering new opportunities for game design and future research in AR-driven passenger entertainment. Von Sawitzky et al. \cite{Sawitzky23ArPlacement} investigated how passengers in fully automated vehicles might position infotainment content using an AR interface, rather than relying on traditional windshield displays. Findings from a Virtual Reality (VR) user study indicate that passengers tend to place content in non-obstructive areas, such as the dashboard. This reflects a desire to remain informed about the driving process.

However, challenges emerge when employing immersive technologies in transit. McGill et al. \cite{mcgill2020challenges} investigated the use of AR and VR HMDs in such contexts, highlighting their potential to enhance passenger productivity, privacy, and immersion. Their findings reveal significant barriers, including motion sickness, crash safety concerns, and social acceptability. Similarly, Togwell et al. \cite{Togwell2022gaming} also highlight challenges for in-car AR. These include the changing external environment, motion sickness, alignment accuracy, and pedestrian privacy concerns. These barriers must be addressed to fully realize the benefits of AR headsets in transit environments, underscoring the need for further research in this area.

In summary, while NDRTs for passengers have been widely studied and in-car AR has been explored in certain areas, significant challenges and gaps remain. Our research addresses these by examining passengers' habits and preferences for interacting with location-based data. Furthermore, we present and evaluate a novel AR-based infotainment system that enables passengers to interact with POIs through an intuitive interface, an approach not previously explored.

\subsection{Interactions for In-Car Augmented Reality}
Selection is a critical task in interactions between users and virtual elements in AR systems \cite{blattgerste2018advantages, Doerner2022}. Although object selection in AR has been extensively studied \cite{nizam2018review, hertel2021taxonomy, kyto2018pinpointing, zhou2008trends, blattgerste2018advantages}, most research has focused on stationary environments. Nonetheless, insights from these studies can inform human-computer interaction research in dynamic settings, such as moving vehicles. 
Performing mid-air interactions in moving vehicles presents unique challenges primarily due to the unpredictable nature of vehicular motion. Previous research has explored various aspects of interaction within these dynamic environments, focusing on the potential of multimodal input methods to enhance usability during vehicular motion \cite{roider2018SeeYourPoint}. These approaches aim to address the disruptions caused by vehicle movement, offering more reliable and safe UIs in moving vehicles.

Related work has demonstrated that eye-gaze is often the preferred interaction technique among users and consistently performs better than alternatives. For example, Blattgerste et al. \cite{blattgerste2018advantages} found that eye-gaze outperforms head gaze in several parameters, including speed, while Luro and Sundstedt \cite{luro2019comparative} noted that eye-gaze can reduce cognitive load in participants. Hansen et al. \cite{hansen2018fitts} found that while eye-gaze had lower accuracy and throughput compared to head gaze, the latter was more physically demanding. Kyt{\"o} et al. \cite{kyto2018pinpointing} further showed that although eye-gaze is generally faster, head pointing tends to be more accurate. However, these works were limited to stationary environments only.

There are existing works that studied the selection of objects from the inside of a vehicle. However, previous works mostly study interaction with objects inside the car \cite{aftab2020point}, during short driving rounds \cite{gomaa2020studying}, or don't use AR. For example, R{\"u}melin et al. \cite{rumelin2013free} and Fujimura et al. \cite{fujimura2013driver} investigated hand pointing for interaction with distant objects while in the vehicle, but did not use any immersive technologies. Aftab et al. \cite{aftab2021multimodal, aftab2022pointingoutside} explored multimodal interaction of head, eye, and finger direction for drivers referencing outside-vehicle objects. Gomaa et al. \cite{gomaa2020studying} also adopted a multimodal approach to use eye-gaze and hand pointing to reference objects outside the car. While hand-based interaction has been commonly studied inside vehicles, McGill et al. \cite{mcgill2020challenges} pointed out that physical constraints of the in-car environment, such as available space and motion restraints, may impair the effectiveness of such techniques.

Furthermore, only limited research has focused on interactions performed within AR in vehicles \cite{Schramm2023Assessing, colley2022swivr, kari2023handycast, Tseng2023FingerMapper}. Studies by Tseng et al. \cite{Tseng2023FingerMapper} and Kari et al. \cite{kari2023handycast} have explored interaction techniques that could improve usability within the confined spaces of cars. Colley et al. \cite{colley2022swivr} used a one-degree-of-freedom motion platform to examine standard input methods in VR, assessing their effectiveness in terms of task performance under vehicular motion. The findings of Schramm et al. \cite{Schramm2023Assessing} indicated that eye-gaze with a hardware button achieved the highest selection speed and lowest error rate. However, these results were specific to seat-fixed elements, prompting further investigation into the suitability of eye-gaze techniques for interacting with world-fixed elements.

In summary, multiple interaction techniques have been widely studied for in-car interaction, including hand-pointing, eye-gaze, head-gaze, and multimodal approaches. However, most of these studies focused on interaction without AR and on objects within the vehicle. Building on the demonstrated preference for eye-gaze in previous works \cite{blattgerste2018advantages, luro2019comparative, Schramm2023Assessing}, our research contributes to the field by evaluating eye-based interaction in AR for both car-fixed UIs and world-fixed POIs, expanding the scope of interaction techniques to include external, AR-enhanced environments.

\subsection{Location-based Data}
There is limited research exploring interaction with a cars' outside environment, especially in regard to passenger-based AR \cite{Berger21InteractiveCarDoor}. Matsumura and Kirk \cite{MatsumuraActivePassengering18} investigated the potential of interactive car window systems to enhance passenger engagement with the external environment during car journeys. Their research identifies key themes that support an improved passenger experience, such as:
\begin{itemize}
    \item Active Participation: Passengers appreciated the system for providing a more defined role during the journey, allowing them to engage more actively.
    \item Reflective Interaction: Participants expressed a desire to use the system for post-journey reflection, adding a layer of meaning to their travel experiences.
    \item Social Connectivity: The system's potential to foster social interactions within the car was highlighted, emphasizing its role in enhancing in-car social dynamics.
    \item Temporal Awareness: The system prompted reflections on the changing nature of time during the journey, influencing how passengers perceive and engage with the passage of time.
\end{itemize}

Regarding AR-based POI exploration, Berger et al. \cite{Berger21InteractiveCarDoor} presented an interactive car door concept for rear-seat passengers, featuring an AR-enabled side window that displays POIs along a route, with additional information accessible via a touch-sensitive door panel. Their remote pilot study demonstrated that participants found this system enhanced their user experience, making the ride more engaging and informative by providing trip progress updates and detailed POI information. Overall, the concept was well-received, with participants indicating it made the rear-seat experience more attractive and enjoyable.

While no scientific source exists specifically for an AR concept for interaction with POIs, related concepts have been explored by car manufacturers. For example, BMW presented a concept involving multimodal interaction with the vehicle and its surroundings\footnote{BMW Group: Natural and fully multimodal interaction with the vehicle and its surroundings. \url{https://www.press.bmwgroup.com/global/article/detail/T0292196EN/natural-and-fully-multimodal-interaction-with-the-vehicle-and-its-surroundings-bmw-group-presents-bmw-natural-interaction-for-the-first-time-at-mobile-world-congress-2019} (accessed on 12.11.2024)}\textsuperscript{,}\footnote{BMW Group: BMW Natural Interaction. \url{https://www.press.bmwgroup.com/global/video/detail/PF0006701/bmw-natural-interaction} (accessed on 12.11.2024)}. In this concept, a gesture pointing at a building allows the driver to receive additional information about the building. The work by Aftab et al. \cite{aftab2020point} is directly motivated by this concept, presenting the technical implementation for the multimodal gesture detection. However, no studies exploring how people use this system have been performed.

In summary, while some concepts explore interaction with a car's external environment, none have examined the use of AR for interacting with virtual objects in that environment. We address this gap by introducing a system that leverages eye-gaze and hand gestures, enabling users to interact with POIs in their surroundings. Additionally, our system provides detailed information about the POIs and allows users to explore both upcoming and passed POIs.
\section{Survey on current Passenger behavior regarding POIs}
\label{sec:survey}
In this Section, we outline our survey to assess passengers needs for interaction with their environment while in transit. The survey consisted of structured questions related to individuals' driving habits and their preferences regarding POIs as both drivers and passengers. The questionnaire included multiple-choice questions, Likert-scale items, and two open-ended questions. It was divided into three main Sections: demographics, general questions, and specific scenarios for passengers and drivers. The overarching research question for the survey was:
\begin{itemize}
    \item \textbf{RQ1$_{survey}$:} What informations do passengers need to successfully find passed POIs while on the move?
    \item \textbf{RQ2$_{survey}$:} In what ways do passengers want to interact with POIs while on the move?
\end{itemize}

\subsection{Survey Design}
Participants were asked about the frequency of their driving, with options ranging from \textit{daily} to \textit{less than once a month}, and the frequency of being a passenger, using the same set of options. They were also asked if they had ever created a list of places to visit, followed by questions about the tools used for creating such lists and for navigation purposes. Multiple selections were allowed. To ensure relevance, participants were asked about the frequency with which they assumed the roles of either driver or passenger. They could only fill out the questionnaires for which role they are familar with by not selecting \textit{never} as a frequency.

Passenger-specific questions focused on individuals' experiences as passengers, including their involvement in navigation, the tools they use, and their preferences for saving and recalling locations of interest encountered during travel. The survey asked passengers about the frequency of assisting drivers with directions, their use of navigation tools, and their interest in features for saving and retrieving information about places observed during the journey.

Driver-specific questions examined drivers' perspectives, including how they plan routes, utilize navigation tools, and manage the discovery of new places while driving. As with the passenger section, drivers were asked about their interest in features that enable the saving of POIs, either manually or automatically. The survey also investigated the challenges faced by both drivers and passengers in recalling the names of places they passed, as well as the types of information needed to facilitate later identification of these locations. Finally, participants were asked to express their frustrations with current navigation tools, with the goal of identifying features to avoid when designing new tools.

\subsection{Respondents Demographics}
We conducted the survey with employees of an automotive software consulting company. A total of 110 individuals responded, comprising 81 males, 28 females, and 1 individual who preferred not to specify their gender. The respondents' ages ranged from 23 to 63 years ($mean = 39, SD = 9.61$). Three participants (2.7\%) do not drive regularly and, therefore, did not complete the driver-specific questionnaire, resulting in 107 valid responses. Among these drivers, 37\% drive daily, while 43\% drive three to four times per week. In terms of navigation system usage, 35\% utilize such systems occasionally, and 37\% frequently, encompassing both built-in systems and smartphone-based applications. Additionally, 14\% report using navigation systems every time they drive. Four participants (3.6\%) are not regular passengers. As a result, they did not complete the passenger questionnaire, leaving 106 valid responses for the passenger-specific survey. Among these respondents, 33\% reported being passengers infrequently (1-3 times per month), while 32\% indicated they are passengers once a week.

\subsection{Survey Insights}
\label{surveyResults}
As the survey included both multiple-choice questions and free-text fields, we analyzed and reported the frequencies of responses for each relevant question, distinguishing between passengers and drivers. The most insightful findings resulted from the questions concerning missed POIs and the process of saving POIs.

\subsubsection*{\textbf{Missed POIs}}
\begin{figure}[ht]
    \centering
    \includegraphics[width=\linewidth]{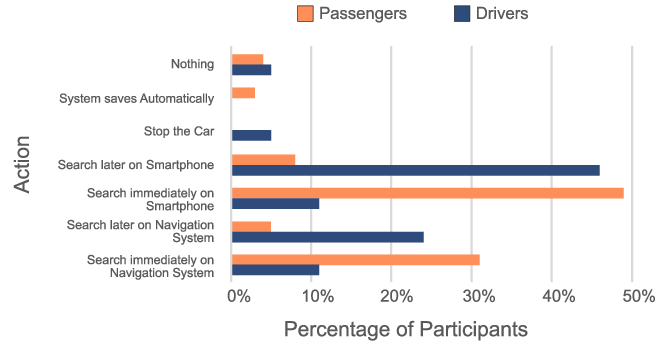}
    \caption{A Graph showing how and when passengers and drivers look for a missed point of interest. Multiple choice was possible.}
    \label{fig:survey_search_behaviour}
    \Description{Barcharts showing participants' behavior when missing a point of interest, grouped by drivers and passengers. Bars represent the percentage of users selecting the multiple-choice answers. Mean values are provided in the appendix. 'Later on smartphone' has the most votes by drivers, whereas 'search immediately on smartphone' was picked by most passengers.}
\end{figure}

Our survey results indicate that the majority of passengers experience difficulty in recalling the names of missed locations they are interested in, with 64.2\% reporting occasional challenges and 4.7\% always facing this issue. This tendency is even more pronounced among drivers, likely due to the demands of focusing on the driving task; 77.6\% of drivers report sometimes struggling to remember POI names, while 5.6\% consistently encounter this difficulty. Figure \ref{fig:survey_search_behaviour} illustrates the percentages of individuals' responses when they pass by an interesting location and miss it. Notably, 49.1\% of passengers immediately search for the location on the web using their smartphones, while 31.1\% conduct a similar search using the vehicle's navigation system. This behavior aligns with related literature on NDRTs \cite{russell2011passengers, hecht2020ndrts, MatsumuraActivePassengering18, BergerGridStudyInCarPassenger2021} and underscores the importance of environmental interaction for passengers. Regarding the importance of specific information for recognizing a POI, there was significant consensus (90\%) on the necessity of knowing the POI's name to accomplish this task for both drivers and passengers. However, passengers placed additional emphasis on the need to know the \textit{category} and have a \textit{description} of the place to recognize it, compared to drivers.

\subsubsection*{\textbf{Saving POIs}}
In the survey, 74.8\% of drivers and 67.9\% of passengers expressed a desire for the functionality to save POIs in their navigation systems, indicating a significant interest in this feature across both groups. This interest corresponds with the broader goal of enhancing user engagement with their environment, particularly in relation to the creation of POI lists. Regarding the saving of POIs, 65.5\% (N = 72) of participants reported having created a list of POIs at least once in their lifetime. Both groups demonstrated a clear preference for manual saving, with 93.1\% of passengers and 65\% of drivers favoring this method. Additionally, 18.8\% of drivers preferred automatic saving, which is understandable given the demands of driving. Therefore, an in-car POI system should be designed to accommodate the needs of both passengers and drivers, offering the option to save POIs with a strong emphasis on manual saving.
\section{Pre-Study on Eye-Gaze Interaction}
\label{sec:pre-study}
In a pre-study with 10 participants, we investigated the feasibility of interacting with world-fixed content as a passenger using AR in a moving vehicle. We employed the technique of using eye-gaze to hover over POIs and confirming the selection via a hardware button, as this was shown in \cite{Schramm2023Assessing} to be one of the favored selection techniques for interacting with in-car AR content. The key distinction between our study and that of Schramm et al. \cite{Schramm2023Assessing} is that our POIs are world-fixed, rather than car-fixed. As a result, the task structure in our study differed, as outlined in Section \ref{sec:prestudy_procedure}. The time available for participants to select a POI in our study was thus not fixed but instead varied based on current traffic conditions, providing a closer scenario to real-world conditions. The research question guiding our pre-study was as follows:

\begin{itemize}
    \item \textbf{RQ$_{pre}$:} Is interaction with world-fixed content via eye-gaze combined with a hardware button a feasible method for in-car AR?
\end{itemize}

\subsection{Participants and Apparatus}
\label{sec:pre-study_apparatus}
\begin{figure*}[ht]
    \centering
    \includegraphics[width=\linewidth]{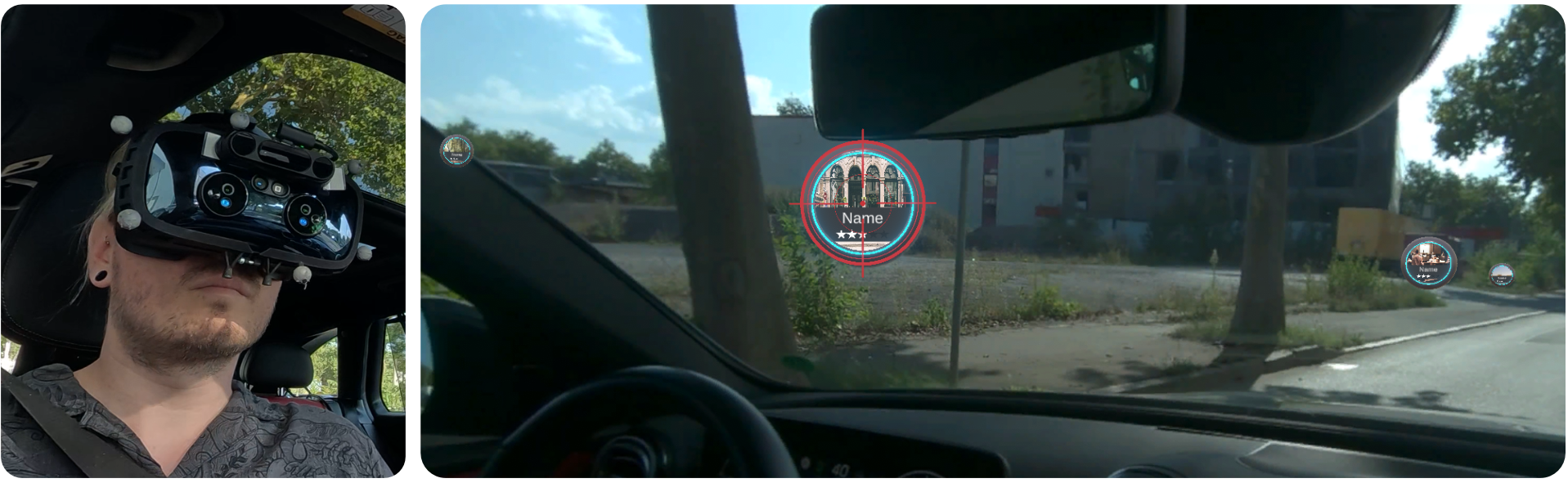}
    \caption{View of the pre-study. We use the Varjo XR-3 with additional optical tracking (left). POIs are visualized as spheres outside the vehicle (right). The POI with the red crosshair had to be selected via eye-gaze and a hardware button.}
    \label{fig:prestudy_varjoview}
    \Description{The setup for our pre-study. On the left, a person sitting in a car, wearing the Varjo XR-3 headset. On the right image, the view inside the Varjo XR-3 is displayed, showing the scene through the car's windshield. Outside the car are four spheres representing points of interest. The closest point of interest is marked with a red crosshair.}
\end{figure*}

We recruited ten participants, comprising two females and eight males, with a mean age of 32.5 years ($SD = 7.17$). Participants were seated in the front passenger seat of a premium midsize estate vehicle. The study was conducted on a street in a public industrial area to simulate realistic driving conditions. The chosen track, labeled as \textit{Pre-study track} in Figure \ref{fig:study_tracks}, was around 3.5 kilometers long, had a 50 km/h speed limit, and featured wide roads with moderate mixed traffic, including cars, busses, trucks, bicycles, and pedestrians. Some road sections were bumpy due to frequent heavy vehicle traffic. To ensure uniform driving conditions, the car's speed limiter was set to the maximum allowable speed of 50 km/h. Fourty POIs were placed around the track. Each of them was located perpendicular to the center of the street at a distance of 7.5 meters, alternating to the left and to the right side of the street. The POIs had a diameter of three meters and showed fake restaurants comprised of fictional names and images. The pre-study setup is shown in Figure \ref{fig:prestudy_varjoview}.

For the AR hardware, we selected the Varjo XR-3\footnote{\label{foot:Varjo}Varjo Technologies Oy: Varjo XR-3, the first true mixed reality headset. \url{https://varjo.com/products/varjo-xr-3/} (accessed on 26.08.2024)} video see-through (VST) HMD, chosen for its advanced features and compatibility with the middleware from LP-Research\footnote{\label{foot:lpvr}LPVR Middleware a Full Solution for AR / VR. \url{https://www.lp-research.com/middleware-full-solution-ar-vr/} (accessed on 12.09.2024)}. This setup enabled 6-Degrees of Freedom (DoF) HMD tracking in a moving vehicle, supported by an additional car-mounted inertial measurement unit. The selection technique was implemented using Microsoft's Mixed Reality Toolkit (MRTK) Version 2.8.3\footnote{Mixed Reality Toolkit 2. \url{https://learn.microsoft.com/en-us/windows/mixed-reality/mrtk-unity/mrtk2/} (accessed on 12.09.2024)} in Unity.

\subsection{Procedure and Task}
\label{sec:prestudy_procedure}
Participants first completed a declaration of consent and a demographic questionnaire. Following this, they were seated in the car and were provided with an explanation of the procedure and the task. Participants were instructed to only select POIs marked with a crosshair (as seen in Figure \ref{fig:prestudy_varjoview}) and to do so as fast as they could, following a procedure similar to that used in \cite{Schramm2023Assessing}. In our study, the crosshair was randomly placed on a single POI located within a radius of 70 meters in front of the vehicle. If the participant successfully selected the marked POI or the vehicle passed the marked POI, another nearby POI was randomly marked. Each marked POI was on average marked for 4.80s (Mdn = 2.87s, SD = 6.32s) before being selected or passing the car. Each round lasted between five and six minutes, depending on the traffic conditions on the study track. After completing the task, participants filled out the Raw NASA Task Load Index (RTLX) \cite{hart1988development, hart2006nasa} and System Usability Scale (SUS) \cite{Brooke96SUS} questionnaires. This was followed by a short semi-structured interview to gather qualitative feedback on participants' preferences.

\subsection{Pre-Study Results and Discussion}
\label{sec:prestudy_results}
\subsubsection*{\textbf{Error Rate:}} Among the 724 marked POIs across all participants, 483 (66.71\%) were correctly selected, while 170 (23.48\%) were missed.  Additionally, for 71 (9.81\%) marked POIs, an unmarked POI was incorrectly selected instead. These results are less favorable compared to the findings of Schramm et al. \cite{Schramm2023Assessing}, where in the eye-tracking condition, 7.79\% of marked elements were missed, and 2.86\% of unmarked elements were erroneously selected. The discrepancy likely stens from the differences in the placement of POIs. Unlike the car-fixed POIs used in the study by Schramm et al. \cite{Schramm2023Assessing}, our world-fixed POIs represent moving targets, making them more susceptible to being missed.

\subsubsection*{\textbf{Task Completion Time:}} We also measured the time elapsed between marking a POI and subsequently selecting the marked POI. The mean time for selection was 1.82 seconds (Mdn = 1.35s, SD = 1.50s). This result closely aligns with those reported by Schramm et al. \cite{Schramm2023Assessing}, where the mean time for selection using eye-gaze with hardware confirmation was also 1.82 seconds (Mdn = 1.54s, SD = 0.913). While our median time is 0.19 seconds shorter, our standard deviation is 0.587 seconds higher. These findings suggest that the placement of world-fixed POIs, as compared to car-fixed POIs, does not have a significant effect on the time required to select a marked POI in our scenario.

\subsubsection*{\textbf{Perceived Workload}:} Our system received a mean workload score of 24.8, which is in line with related literature. In the work of Kyt{\"o} et al. \cite{kyto2018pinpointing}, interacting via eye + device while standing had a similar mean RTLX score of roughly 30. Though, comparability is limited, as their tasks took significantly longer and they used a Microsoft Hololens for testing. Schramm et al. \cite{Schramm2023Assessing} evaluated the same technique with a similar in-car setup also using the Varjo XR-3. Their mean RTLX score for eye + hardware confirmation of 24.6 is close to ours. Blattgerste et al. \cite{blattgerste2018advantages} also received similar values to us for RTLX using eye-gaze while being stationary. They evaluated the workload for three Fields of View (FOVs), where the large (90\textdegree{}) FOV is closest to the 110\textdegree{} FOV of the Varjo XR-3. In this condition, they measured a RTLX value of 27.5, which is also close to ours. To summarize, both Kyt{\"o} et al. \cite{kyto2018pinpointing} and Schramm et al. \cite{Schramm2023Assessing} conclude that eyegaze + device are feasible selection methods  with similar RTLX scores to ours. In addition, \cite{blattgerste2018advantages} shows the advantages of eye-gaze, also with similar RTLX to ours. Thus we conclude that this technique is also feasible for in-car AR use regarding workload. 

\subsubsection*{\textbf{Usability:}} Our system received a mean SUS score of 86.0 (Mdn = 87.5, SD = 8.01), which corresponds to an \textit{excellent} rating according to Bangor et al. \cite{bangor2009sus}. This score is consistent with related literature, as Schramm et al. \cite{Schramm2023Assessing} achieved a similar SUS score of 85.6 for their eye-gaze and hardware condition. Thus, we can conclude for RQ$_{pre}$ that eye-gaze combined with a hardware button is a feasible method for interacting with world-fixed objects in a moving vehicle.

\begin{figure*}[ht]
    \centering
    \includegraphics[width=\linewidth]{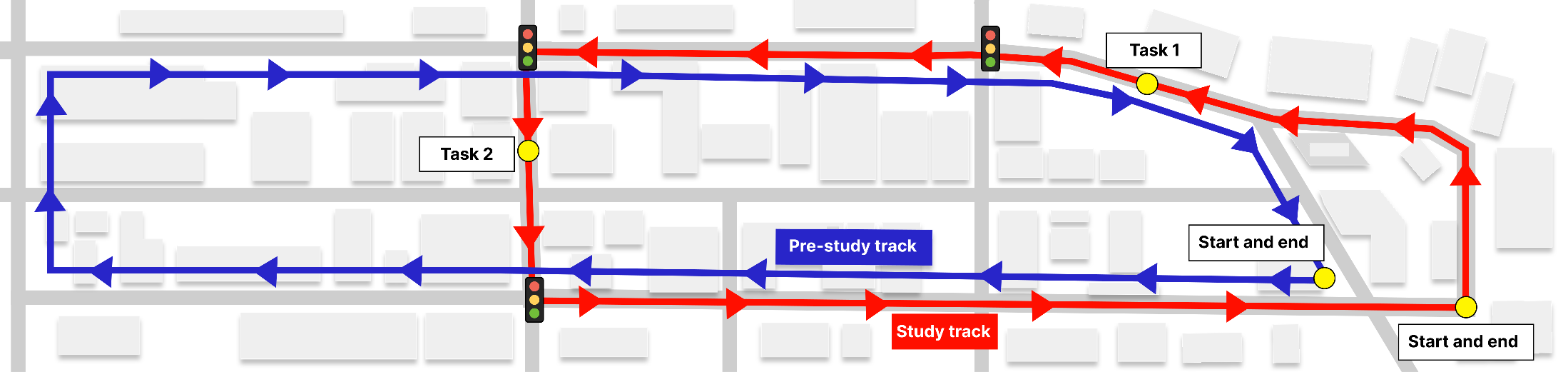}
    \caption{The tracks used for the pre-study (blue) and the main-study (red). The studies were conducted in an industrial area with a 50km/h speed limit and moderate traffic. Traffic lights are annotated via icons.}
    \label{fig:study_tracks}
    \Description{A top-down 2D map with two driving tracks marked in different colors, representing the pre-study track and the study track. The pre-study track is highlighted in blue and forms a rectangular loop across two and a half blocks. The study track is marked in red and forms a similar loop. Both tracks start and end at roughly the same location, indicated by a yellow circle labeled "Start and end." For the study-track, two tasks are marked on the map. Task one occurs at the 20\% mark of the track. Task two is located at 60\% of the track. Traffic lights are shown at two intersections along the shared route, which both tracks pass through. Arrows on each track indicate the direction of travel for both routes, with the pre-study track going clockwise and the study track going counterclockwise.}
\end{figure*}

\section{Study on Visualizing passed and upcoming POIs}
\label{section:study}
We created and evaluated a prototype to test a realistic scenario where a passenger can use AR to explore their environment through digital POIs overlaid onto the real world. The study employed a within-subjects design and was conducted in a car setting in the field. The research questions guiding our investigation are outlined below:
\begin{itemize}
    \item \textbf{RQ1$_{main}$:} How can AR effectively support passengers in discovering missed and upcoming POIs?
    \item \textbf{RQ2$_{main}$:} Is eye-gaze and pinch a feasible interaction method for interacting with both world-fixed POIs and car-fixed UIs?
    \item \textbf{RQ3$_{main}$:} Would users accept an VST-based in-car AR system to explore POIs?
\end{itemize}

\begin{figure*}[ht]
    \centering
    \includegraphics[width=\linewidth]{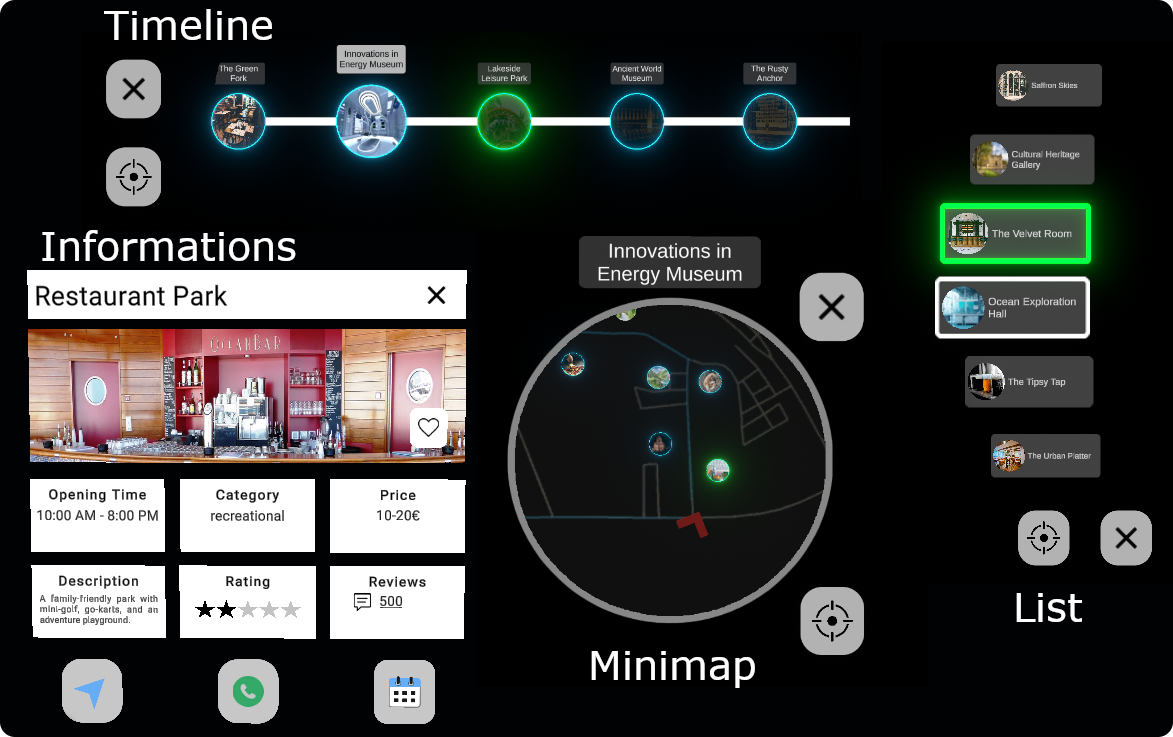}
    \caption{An overview of the UI elements used in the study. The \textit{Informations} panel always correspondet to the currently selected POI. The \textit{List}, \textit{Minimap}, and \textit{Timeline} were only used during their study-conditions respectively.}
    \label{fig:study_design_elements}
    \Description{The image depicts an interactive user interface for a navigation system with several key sections: Timeline (top center) shows a horizontal sequence of locations with circular icons representing different points of interest (POIs). 'Informations' (bottom left) displays details for a single POI, including opening times, category, price, a description, rating, and reviews. A photo of the location is also included. Minimap (center right) highlights POIs in a small, circular map with selectable icons. List (right side) provides a vertical list of locations, with one POI highlighted in green. Control buttons ("X" to close and a crosshair to re-center) are present in each section for navigation and interaction.}
\end{figure*}

\subsection{Prototype Design}
\label{sec:study_prototype_design}
Our prototype design is grounded in the findings from our survey (Section \ref{sec:survey}) and pre-study (Section \ref{sec:pre-study}). The survey results indicated that passengers often experience difficulty recalling the names of missed POIs and prefer to search immediately for interesting locations on the web using their phones. Additionally, many respondents expressed a desire for the functionality to save POIs. Therefore, the primary focus of our study is to investigate methods to increase the success rate of participants in rediscovering POIs and to interact with those that are not in the vehicle's immediate surroundings. In addition to the name, we include an image, a category, and a description for each location to assist users in easily identifying it. We include three possible categories for POIs, with each category including three types for variation: food (restaurants, bars, and pizza), museums (art, science, and history), and parks (zoos, nature, and recreational). The system adheres to Nielsen's sixth heuristic principle, which prioritizes recognition over recall \cite{Nielsen1994Usability}. These and more informations are visible after selecting a POI, as shown in Figure \ref{fig:study_design_elements} at \textit{Informations}.

To explore passed and upcoming POIs, we designed three in-vehicle visualizations to be used in tandem with the world-fixed POIs outside the vehicle: \textit{Timeline}, \textit{List}, and \textit{Minimap}. The designs are illustrated in Figure \ref{fig:study_design_elements}. The system was designed to minimize cognitive load by enabling users to recognize places within a sequence rather than recall specific names \cite{Nielsen1994Usability}. The \textit{Timeline} is designed to emphasize chronology, reflecting the sequential nature of encountering POIs along a route. The \textit{List}, while conveying order, does not inherently suggest a chronological sequence, as lists can be organized in various ways, such as alphabetically or sorted by rating. Consequently, we anticipated that participants might interpret the location of past and future POIs in the \textit{List} differently from each other. The \textit{Minimap} was expected to convey a sense of spatial chronology, as POIs would appear sequentially along the route. All visualizations were designed to avoid blocking the outside view, as found by Sawitzky et al. \cite{Sawitzky23ArPlacement}, and were present in the FOV only when the user actively engaged with them.

Regarding graphical choices, the timeline was designed to allow the user to constantly track progress, with each POI represented by a dot along the route. To express progress, the timeline fills in as the next POI approaches. The list displays the POI closest to the user as the central element, with future POIs located below it and past POIs above. To facilitate progress tracking, the list automatically moves to the current POI when opened. The map displays only the segment of the route currently being traversed, with POIs indicated by dots. Progress is tracked by a moving arrow that represents the cars position. Each of the three visualizations also includes a button allowing the user to quickly return to the current position and a button to close it.

\begin{figure}[ht]
    \centering
    \includegraphics[width=0.6\linewidth, trim={0.5cm 1cm 0.5cm 0.5cm},clip]{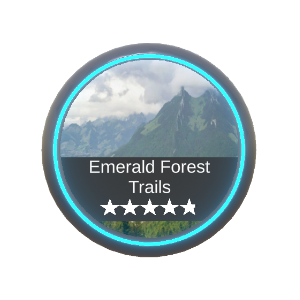}
    \caption{Design of the world-fixed POIs with the location name, star rating, and representative image.}
    \label{fig:study_design_POI}
    \Description{A circle featuring a grey outer border, with a smaller inner border that resembles a glowing neon tube, emitting a light blue light. The circle showcases an image of a mountain scenery. It also includes a dark grey bar across the bottom of the image, displaying the name and star-rating of the respective point of interest.}
\end{figure}

Similar to the pre-study apparatus in Section \ref{sec:pre-study_apparatus}, we placed fourty POIs around the study track. Each of them was located perpendicular to the center of the street at a distance between 5 and 7.5 meters, alternating to the left and to the right side of the street. The POIs had a diameter of three meters and showed fake locations comprised of fictional names and images. Location categories included food, mueums, and parks. This setup simulated the experience of exploring a new city route, with POIs unknown to the participants, helping them find new places to visit during the trip. World-fixed POIs followed the design shown in Figure  \ref{fig:study_design_POI}. 

We used eye-gaze as a pre-selection method for both world-fixed and car-fixed content, based on the findings from our pre-study in Section \ref{sec:pre-study} and the findings by Schramm et al. \cite{Schramm2023Assessing}. Since the car-fixed visualizations require scrolling, we integrated hand-tracking techniques. As such, Eye-gaze is used to preselect elements for interaction, and pinching is used to confirm the selection. Scrolling works by holding the pinch gesture and Simultaneously moving the hand in the desired direction. We limited interactions to the participants' dominant hand to mitigate accidental selections through the non-dominant hand.

\subsection{Participants and Apparatus}
\label{sec:study_participants_apparatus}
We conducted the study with employees of an automotive software consulting company, recruited through convenience sampling via email invitation and word of mouth (N = 21; 6 female, 15 male; mean age = 36.0 years, SD = 11.4 years). Participants were asked to rate their experience with immersive technologies using three 5-point Likert scales, ranging from no experience to extensive experience. The three categories assessed were experience with HMDs (M = 2.62, Mdn = 2, SD = 1.47), interaction via eye-tracking (M = 1.90, Mdn = 1, SD = 1.18), and hand-tracking (M = 2.29, Mdn = 2, SD = 1.35). Additionally, twelve participants required corrective eyewear, while nine did not. Seventeen participants were right-handed, four were left-handed.

Similar to the pre-study apparatus described in Section \ref{sec:pre-study_apparatus}, we used the Varjo XR-3\footref{foot:Varjo} VST HMD with LP-Research\footref{foot:lpvr} 6-DoF tracking. We attached a Leap Motion Controller 2\footnote{\label{foot:Leapmotion2}Ultraleap: Leap Motion Controller 2. \url{https://leap2.ultraleap.com/products/leap-motion-controller-2/} (accessed on 26.08.2024)} to the front of the Varjo XR-3 to have improved handtracking over its' integrated Leap Motion Controller. We used the Unity XR Interaction Toolkit\footnote{Unity Technologies: XR Interaction Toolkit. \url{https://docs.unity3d.com/Packages/com.unity.xr.interaction.toolkit@3.0} (accessed on 12.09.2024)} version 3.0.5 for eye-gaze and pinch interactions.

\subsection{Procedure}
\label{sec:study_procedure}
Initially, participants were asked to sign an informed consent form and to complete a series of demographic questions as reported in Section \ref{sec:study_participants_apparatus}. Then participants were introduced to the study's objectives and procedures. Subsequently, a training phase was conducted within a stationary vehicle to allow participants to familiarize themselves with the eye-gaze and pinching interactions required during the study.

The core experimental phase involved driving participants along a predetermined 3 km track shown in Figure \ref{fig:study_tracks}, once for each condition respectively. As in the pre-study, participants were seated in the front passenger seat of a premium midsize estate vehicle. The car's speed limiter was set to 40 km/h to allow for uniform driving conditions. The lower speed of 40 km/h compared to the 50 km/h in the pre-study is based on the high percentage of missed POIs in the pre-study (Section \ref{sec:prestudy_results}). By lowering the speed, we wanted to mitigate the risk of missing POIs. The mean duration for completing the track was 6.92 minutes (Mdn = 6.74, SD = 0.930).

The three conditions \textit{Timeline}, \textit{Minimap}, and \textit{List} are described in detail in Section \ref{sec:study_prototype_design}. The order of the conditions was counterbalanced using Latin Square to mitigate learning effects. For each condition, participants were asked to complete two tasks, randomly selected from a pool of three options that reflected typical operations passengers might perform while exploring their environment by car. The possible tasks included adding a location to the favorites list, calling a location, and reserving a table or purchasing a ticket depending on the locations category. Participants could complete the tasks by using the buttons on the \textit{Informations} panel, as shown in Figure \ref{fig:study_design_elements}. The tasks were given to the participants by the system at predetermined points on the track to provide participants with sufficient information to perform the tasks. The first task was presented after the vehicle travelled 25\% of the tracks distance, the second tasks was presented after the vehicle travelled 60\% of the tracks distance, as highlighted in Figure \ref{fig:study_tracks}. As the vehicle passed these points on the track, the corresponding tasks was presented to the participant via a pop-up notification, accompanied by a sound cue to capture their attention. Each round consisted of a \textit{past\_task} and a \textit{future\_task}. A \textit{past\_task} relates to a previously seen POI, while a \textit{future\_task} relates to an upcoming POI. The POIs for both task types were randomly chosen and thus were possibly different for each condition to mitigate learning effects. The order of \textit{past\_task} and \textit{future\_task} was also counterbalanced using Latin Square.

Following each condition, participants were required to complete the Motion Sickness Questionnaire (MISC), the RTLX, the User Experience Questionnaire (UEQ), and the SUS. Additionally, a set of custom questions with 5-point Likert scales was administered to collect participants' perceptions of the intuitiveness of the POI display order and the degree to which the display elements may have occluded their view. To further explore the participants' experiences, semi-structured interviews were conducted after each condition and at the conclusion of all conditions. These interviews probed the participants' perceived difficulties, preferences, and suggestions for improvements to the prototype system.

The final phase of the study involved a comparative evaluation, wherein participants were asked to rank the three systems from least to most favorite and to respond to Likert scale questions regarding the perceived usefulness of the system. Throughout the study, participants' responses to questionnaires were recorded using specially designed Excel templates, which facilitated offline completion and automatic updating of a central replies' sheet.

\subsection{Measures}
\label{sec:study_measures}
We employed both quantitative and qualitative measures to evaluate the three presented paradigms for interacting with POIs.

\textbf{Quantitative data.} We first utilized the RTLX \cite{hart1988development, hart2006nasa} to assess participants' perceived workload. Secondly, we employed the SUS \cite{Brooke96SUS} to evaluate the usability of the three proposed solutions. As a third quantitative measure, we assessed user experience using the english 26-item version of the UEQ \cite{Laugwitz2008UEQ}. The UEQ results consist of six factors: attractiveness, perspicuity, efficiency, dependability, stimulation, and novelty. Additionally, we incorporated custom questions after each condition to evaluate participants' interpretation of POI sequencing, the extent to which the interfaces occluded the real world, and their ability to locate POIs based on the tasks. To measure potential motion sickness effects caused by the use of the AR application in a moving vehicle, we employed the MISC \cite{Bos2006Misc}. This scale assessed the severity of motion sickness symptoms, including nausea, dizziness, and headache. The MISC was administered before the study and once after each condition.

\textbf{Qualitative data.} We conducted semi-structured interviews after each condition and at the end of the study. Participants were asked about any difficulties they encountered with the prototype, the aspects they found most challenging to understand, and their most and least liked features. At the conclusion of the study, they were asked to clarify which condition best helped them understand which POIs had been passed and which were upcoming. Additionally, participants were invited to suggest desired features and to discuss when and how they would use the system. Finall, they ranked the three systems from least to most favorite and rated the system's overall usefulness on a scale from 1 to 5.
\section{Results}
\label{section:results}
We conducted a repeated-measures ANOVA with the visualization technique in the study condition as independent variable, and the measures described in Section \ref{sec:study_measures} as dependent variables. When the assumption of normality was violated (tested with the Shapiro-Wilk test), we used the Friedman rank sum test. The post hoc tests were then conducted through pairwise comparisons using the Wilcoxon signed rank test with Bonferroni corrections. All statistical tests are reported at a 0.05 significance level for main effects.

\begin{figure*}[t]
    \centering
    \includegraphics[width=\linewidth]{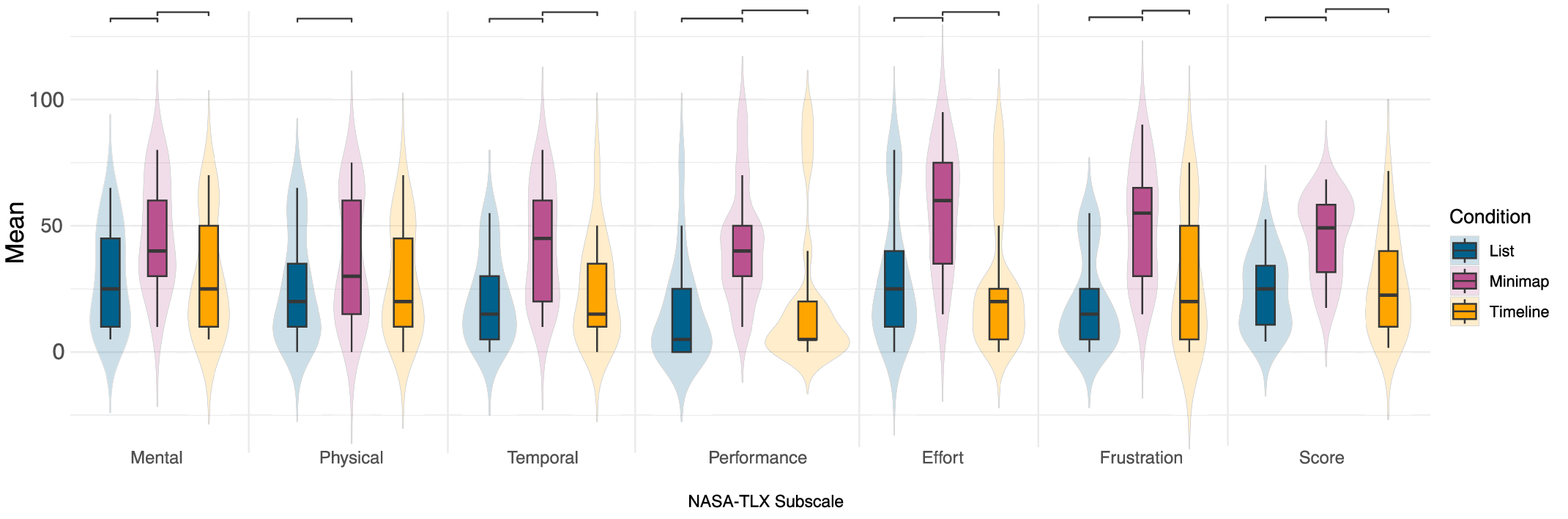}
    \caption{The mean values for each RTLX subscale, split by condition. Scales range from 0 to 100, lower is better. Significant differences are marked with lines between plots.}
    \label{fig:results_RTLX}
    \Description{Violin plots of users' mean NASA TLX workload scores for the three experimental conditions. Mean values are provided in the appendix. The minimap condition scored the highest workload across all conditions. Performance and Effort have few outliers.}
\end{figure*}

\subsection{Workload}
Results for the RTLX questionnaire are presented in Figure \ref{fig:results_RTLX}, with significant differences between conditions indicated. The data did not follow a normal distribution. The outcomes of the repeated-measures ANOVA show, that the minimap condition was associated with a significantly higher task load compared to the other conditions across all categories (Mental: F(2) = 16.48, p < .001; 
Physical: F(2) = 6.86, p = .032;
Temporal: F(2) = 14.29, p < .001; 
Performance: F(2) = 18.81, p < .001; 
Effort: F(2) = 11.08, p = .004; 
Frustration: F(2) = 15.69, p < .001; 
overall TLX-Score: F(2) = 18.28, p < .001).

\begin{figure*}[h]
    \centering
    \begin{subfigure}[b]{.45\textwidth}
        \centering
        \includegraphics[width=\textwidth]{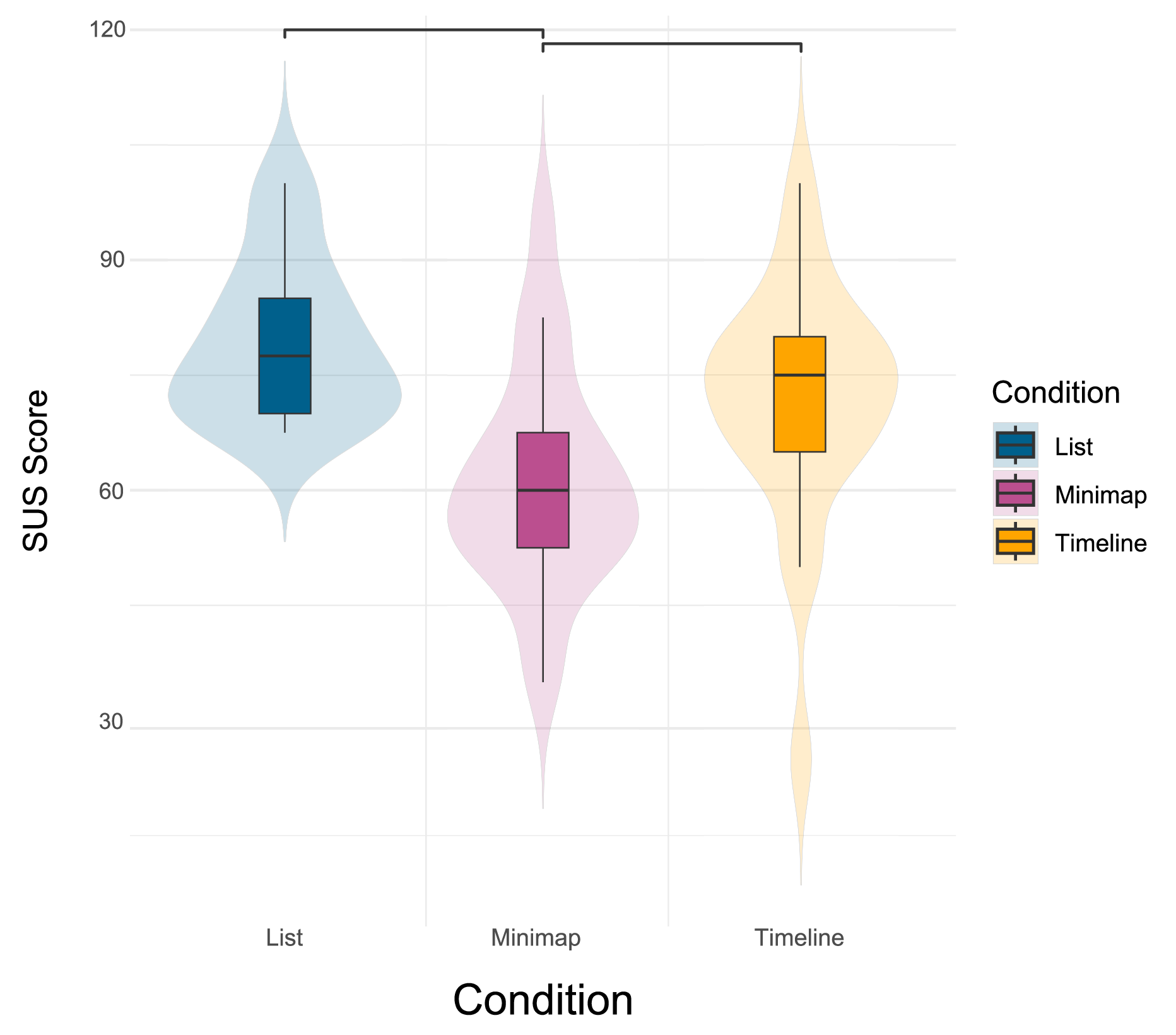}
        \caption{SUS}
        \label{fig:results_SUS}
    \end{subfigure}
    \hfill
    \begin{subfigure}[b]{.45\textwidth}
        \centering
        \includegraphics[width=\textwidth]{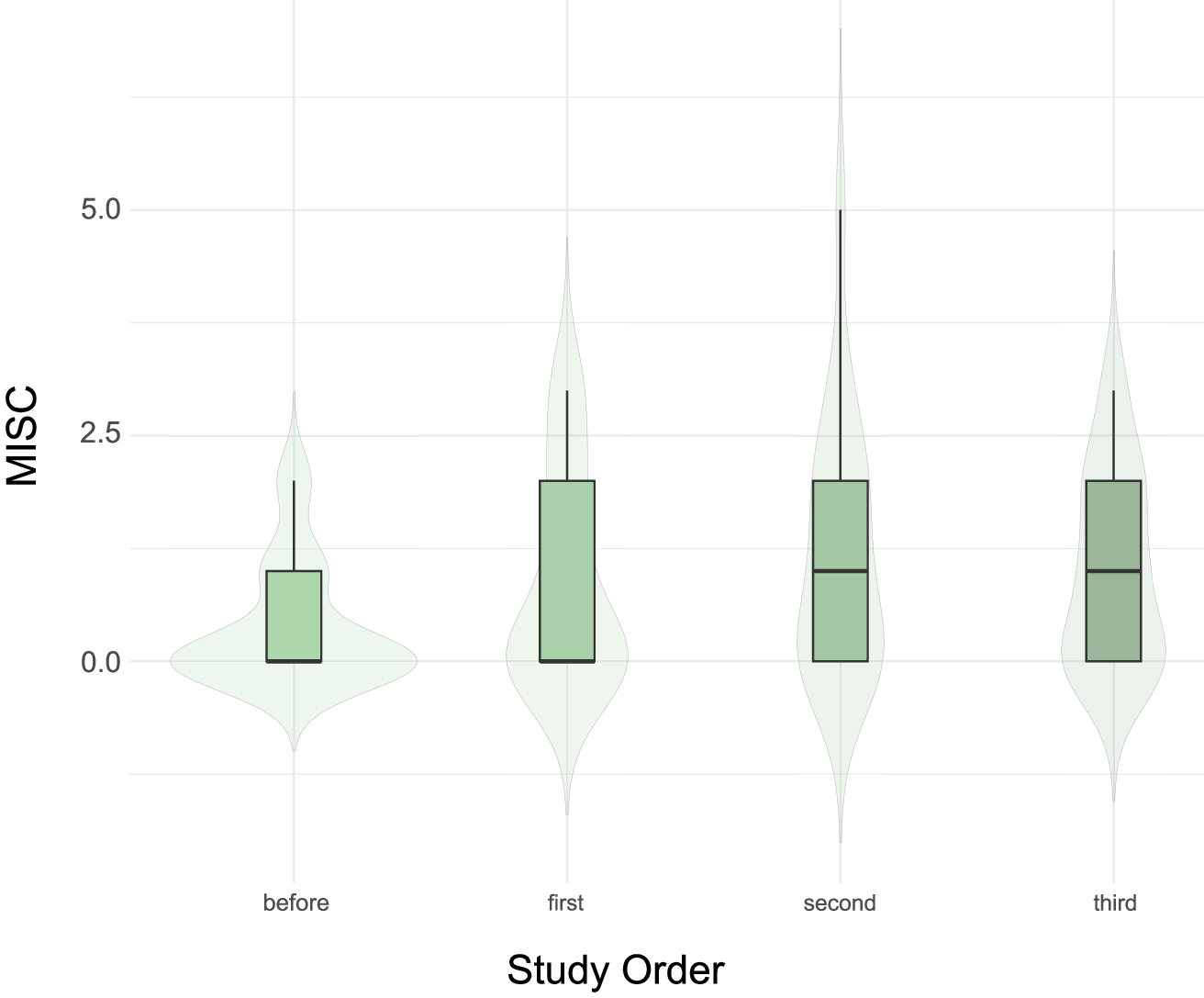}
        \caption{MISC}
        \label{fig:results_MISC}
    \end{subfigure}
    \caption{Left: Violin plots for SUS scores, split by condition. Scales range from 0 to 100, higher is better. Right: Mean MISC scores over the study duration. Scale ranges from 0  to 10, lower is better. Significant differences are marked with lines between plots.}
    \Description{Violin plots of users mean SUS scores (a) and mean MISC scores (b) for the three experimental conditions. Mean values are provided in the appendix. The minimap condition scored a significantly lower usability score than list and timeline. The MISC values show the motion sickness progression over the time the study took place. Scores started between zero and one and rose to scores between zero and two for all three subsequent study runs.}
    \label{fig:results_SUS_MISC}
\end{figure*}

\subsection{Usability}
The results of the SUS questionnaire are presented in Figure \ref{fig:results_SUS}. The distribution of the scores was non-normal. A repeated-measures ANOVA revealed significant differences between conditions (F(2) = 14.46, p < .001). Post hoc tests indicate that the \textit{Minimap} condition had a significantly worse usability than the other two conditions. According to adjective interpretations for SUS scores by Bangor et al. \cite{bangor2009sus}, the  \textit{List} (M = 78.5, Mdn = 77.5, SD = 10.2) and \textit{Timeline} (M = 71.5, Mdn = 75.0, SD = 15.9) conditions can be described having a \textit{good} usability. However, the \textit{Minimap} condition (M = 61.1, Mdn = 60.0, SD = 14.3) could be rated as having an \textit{OK} usability.

\subsection{User Experience}
\begin{figure*}[ht]
    \centering
    \includegraphics[width=\linewidth]{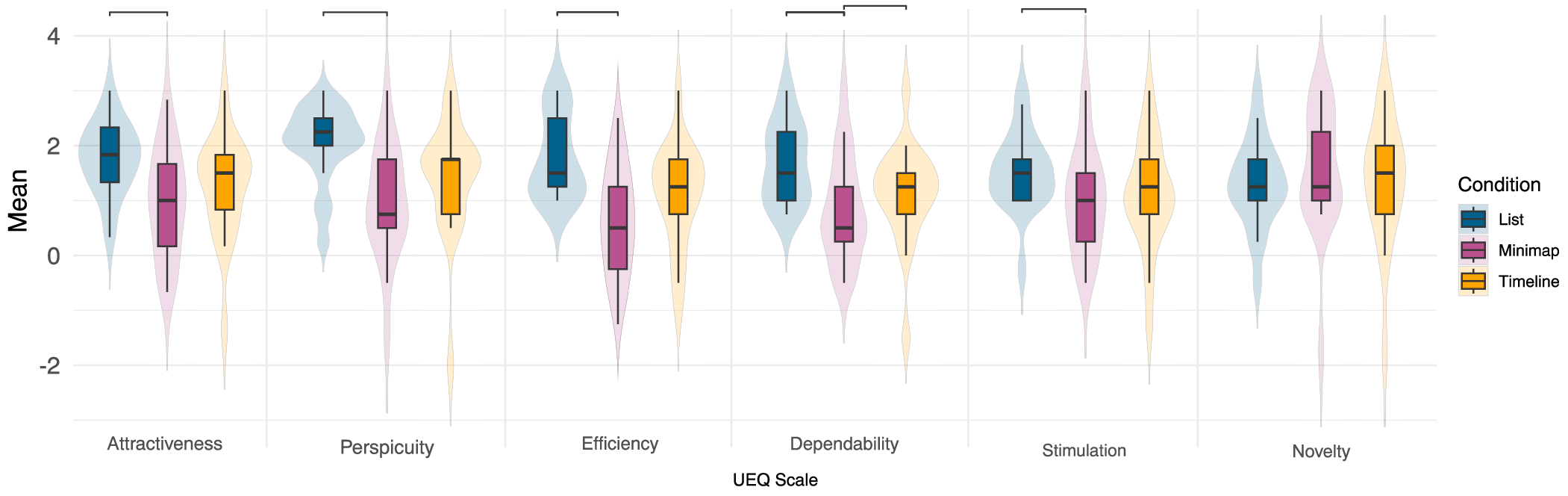}
    \caption{Violin plots for mean values across UEQ scales, split by condition. Scales range from -3 to +3, higher is better. Significant differences are marked with lines between plots.}
    \label{fig:results_UEQ}
    \Description{Violin plots of users mean UEQ scores for the three experimental conditions. Mean values are provided in the appendix. List scored best in Attractiveness, Perspicuity, Efficiency, and Dependability. For Stimulation, List and Timeline have similar scores. For Novelty, all three conditions have similar scores. The minimap has significantly lower scores than list for each subscale except for novelty.}
\end{figure*}

The results of the UEQ questionnaire are presented in Figure \ref{fig:results_UEQ}. The \textit{List} received \textit{excellent} Perspicuity ratings, and \textit{good} ratings for every other scale. The \textit{Timeline} had a \textit{below average} Dependability, \textit{good} Novelty, and \textit{above average} ratings for the remaining scales. \textit{Minimap} had overall \textit{below average} ratings, with only Novelty receiving a \textit{good} score. All scales except Perspicuity showed normal distribution. The Novelty scale violated the assumption of sphercity. Repeated-measures ANOVA showed significant differences between conditions for Attractiveness (F(2) = 7.12, p = .002), Perspicuity (F(2) = 11.60, p = .003), Efficiency (F(2) = 12.04, p < .001), Dependability (F(2) = 10.40, p < .001), and Stimulation (F(2) = 3.47, p = .041). Only Novelty showed no significant differences (F(2) = 2.085, p = 0.35). Pairwise comparisons are shown in Figure \ref{fig:results_UEQ}.

\subsection{Task Completion}
In the \textit{Timeline} and \textit{List} conditions, most participants were able to complete all tasks within the track. The task completion rate was 90.5\% for the \textit{Timeline} and 88.1\% for the \textit{List}. The \textit{Minimap} condition had the lowest task completion rate, with only 59.5\% of the tasks completed within the track.

The time from being given the task and completing it was measured in minutes. The mean task completion time overall was 1.15 minutes (Mdn = 0.88, SD = 0.88). Split by conditions, the \textit{MinimapCondition} took participants the longest (M = 1.67, Mdn = 1.26, SD = 1.24), followed by the \textit{TimelineCondition} (M = 1.04, Mdn = 0.76, SD = 0.74), and the \textit{ListCondition} (M = 0.92, Mdn 0.76, SD = 0.57).

\subsection{Motion Sickness}
Overall, our prototype induced little motion sickness, even though the study took place using a VST HMD in a moving vehicle. The mean baseline score before the study began was 0.38 (Mdn = 0, SD = 0.67), increasing during the study to a mean value of 1.02 (Mdn = 1, SD = 1.18). The mean MISC scores over the time of the study are shown in Figure \ref{fig:results_MISC}. Motion sickness did not increase significantly over the duration of three condition and actually got lower after the third condition again. MISC results also did not significantly differ between conditions, with the \textit{List} having a slightly lower mean impact on MS (M = 0.81, Mdn = 0.93) than the \textit{Minimap} (M = 1.10, Mdn = 1, SD = 1.34) and the \textit{Timeline} (M = 1.14, Mdn = 1, SD = 1.28).

\subsection{Preferences}
\label{sec:results_preferences}
\begin{figure}[ht]
    \centering
    \includegraphics[width=\linewidth]{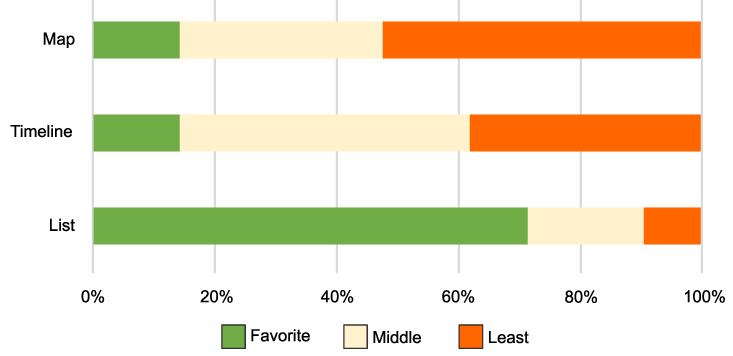}
    \caption{User ranking for the three visualizations investigated in the study. Each participant could rank each visualization from least favourite to favourite.}
    \label{fig:results_preferences}
    \Description{Bar charts showing users' ranking of Minimap, Timeline, and Lists from favorite to least favorite. List is the clear favorite with around 70\% of users voting it as favorite. Minimap and Timeline have similar mixed ratings with minimap having slightly more least favorite votes. Mean values are provided in the appendix.}
\end{figure}

The results of the reported user preferences are presented in Figure \ref{fig:results_preferences}. Among the evaluated visualization methods, the \textit{List} was the most preferred, with 71.4\% (N = 15) of participants selecting it as their top choice. The \textit{Timeline} was most commonly chosen as the middle option, with 47.6\% (N = 10) of participants selecting it in this position. Lastly, the \textit{Minimap} was the least favored, with 52.4\% (N = 11) of participants ranking it as their lowest preference. Participants rated the system's overall usefulness with a mean score of 3.05 (Mdn = 3, SD = 1.02).

Possible influences on the preferences could also stem from the conditions' occlusion of the real world. Participants could rate the conditions occlusion on a 5 point likert scale. For occlusion, the \textit{List} condition was rated with a mean of 2.29 (Mdn = 2, SD = 1.01), the \textit{Minimap} with 2.81 (Mdn = 3, SD = 1.08), and the \textit{Timeline} with 2.86 (Mdn = 3, SD = 0.96). There were no significant differences between conditions.

For the results on how difficult it was for finding the POIs given in the tasks, participants could rate on how easy it was to find on a 5-point likert scale. it was easiest to search in the list with a mean score of 2.00 (Mdn = 2.00, SD = 1.08), similarly easier to search in the timeline (M = 2.24, Mdn = 2.00, SD = 1.14). The map was the most difficult to search in with a mean score of 3.29 (Mdn = 3.50, SD = 0.86). There were significant differences between conditions (F(2) = 23.8, p < .001). Post hoc Tests revealed, that the Minimap was significantly more difficult so search in than the List and in the Timeline.

In addition, participants rated the extent of this occlusion on a 5-point Likert scale. The \textit{List} condition received a mean score of 2.29 (Mdn = 2, SD = 1.01), the \textit{Minimap} a mean of 2.81 (Mdn = 3, SD = 1.08), and the \textit{Timeline} a mean of 2.86 (Mdn = 3, SD = 0.96). No significant differences were found between these conditions.

Participants also rated the difficulty of finding the POIs given in the tasks on a 5-point Likert scale. The \textit{List} was the easiest to search, with a mean score of 2.00 (Mdn = 2.00, SD = 1.08). The \textit{Timeline} was similarly easy to search (M = 2.24, Mdn = 2.00, SD = 1.14). In contrast, the \textit{Minimap} was the most difficult to search, with a mean score of 3.29 (Mdn = 3.50, SD = 0.86). Significant differences were found between conditions (F(2) = 23.8, p < .001). Post hoc tests revealed that the \textit{Minimap} was significantly more difficult to search than both the \textit{List} and the \textit{Timeline}.

For the ordering of POIs in the 2D representations, participants could select one option out of the possible options: chronological, spatial, alphabetical, randomized, and \textit{no idea}. For the list, 47\% of participants understood the chronological order of the List, the rest had no idea or thought this was random. Similary, 42.9\% of participants understood the chronological order of the timeline, whereas 52.4\% of participants understood the spatial order of the Minimap. When prompted, most participants reported, that they did not have the time to really think about the order, as they were preoccupied interacting with a novel system in addition to the tasks.

\subsection{Use-Cases}
During the post-interview, participants were asked when and how they would use the system. Nine participants stated that they would use the system in unfamiliar environments. Their comments included examples such as, "City tours, when you are new in a city," and "In an unfamiliar environment, exploring the surroundings when you are somewhere as a passenger." Within this category, four participants explicitly mentioned using the system while on vacation, with statements such as, "Going on vacation when you want to feel comfortable and discover things.". 

Six participants indicated that they would use the system in their daily lives, providing examples such as, "When i am looking for something specific in everyday life, like a gas station, restaurant, or a specific search with ratings, opening hours, etc.". Additionally, three participants expressed that they would use the system to locate specific destinations, similar to how they would perform a search using a map application on their phone. Regarding locations, six participants explicitly mentioned cities in their responses, whereas two indicated that they would prefer not to use the system on highways.

\section{Discussion}
\label{section:discussion}
In this Section, we discuss and analyze the findings obtained from our user study and semi-structured interview, in relation to the research questions that were previously established.

\subsection{Discovering Passed and Upcoming POIs}
\label{section:discussion_passed_upcoming}
In this section, we answer \textbf{RQ1$_{main}$:} \textit{Which visualization best supports passengers in discovering missed and upcoming POIs?} by utilizing the quantitative data presented in Section \ref{section:results} and the findings from our qualitative interviews.

The overall preferred visualization method in the study was the \textit{List}, as shown in Section \ref{sec:results_preferences} and illustrated in Figure \ref{fig:results_preferences}. Additionally, the \textit{List} had the lowest mean workload scores, alongside the \textit{Timeline}, as shown in Figure \ref{fig:results_RTLX}. The \textit{List} condition also received the highest mean usability score, with an adjective rating of \textit{good} \cite{bangor2009sus}, as seen in Figure \ref{fig:results_SUS}. Among the three conditions, it achieved the highest user experience scores in the categories of attractiveness, perspicuity, efficiency, dependability, and stimulation, as presented in Figure \ref{fig:results_UEQ}. According to the UEQ benchmark \cite{schrepp2017construction}, the \textit{List} condition demonstrated an overall \textit{good} user experience across all scales, with an \textit{excellent} score for perspicuity. The preference for the \textit{List} as the favored visualization technique could be attributed to several factors. One possible reason is the ease of searching within the \textit{List}, as it was rated highest in this category (see Section \ref{sec:results_preferences}), while also causing the least occlusion of the real world. This aligns with comments from some participants (N=6) who noted that the \textit{List} did not overly obstruct their view of the real world and allowed them to still see the POIs in their environment. This finding is consistent with related research, which indicates that passengers generally prefer to maintain a view of their surroundings while traveling \cite{russell2011passengers, hecht2020ndrts, MatsumuraActivePassengering18, BergerGridStudyInCarPassenger2021}. Furthermore, participants appreciated the vertical and familiar layout of the \textit{List}, as most applications for exploring results primarily utilize vertical scrolling \cite{kim2016pagination}.

The \textit{Timeline} visualization received mixed preferences, with most participants ranking it in the middle or last position, as shown in Figure \ref{fig:results_preferences}. However, in some metrics, the \textit{Timeline} performed similarly to the \textit{List}. Its workload was comparably low, and its usability score also fell within the \textit{good} range \cite{bangor2009sus}. In terms of user experience, the \textit{Timeline} was rated lower than the \textit{List}, achieving an overall \textit{above average} rating according to the benchmark \cite{schrepp2017construction}, with the exception of a \textit{good} rating for novelty. For task performance during the study, the \textit{Timeline} performed similarly to the \textit{List}, with no significant differences in task completion rate, task completion time, or ease of searching. Participants also understood the order of the \textit{Timeline} at a rate comparable to the \textit{List}. While occlusion scores for the \textit{Timeline} were higher than those for the \textit{List}, the difference was not significant. Nonetheless, some participants expressed concerns about the \textit{Timeline} occluding too much of the real world. Although there was an option in each condition to close the occluding visualization, it was used sparingly. Interview data revealed that some participants disliked the direction and ordering of the \textit{Timeline}. While all participants understood the purpose of the \textit{List}, the \textit{Timeline} was not immediately clear to everyone. Specifically, the fact that the \textit{Timeline} was intended to represent a chronological sequence was not evident to all participants, leading to confusion about the direction of chronology. This rating may also partially stem from a bias towards vertically scrolling lists \cite{kim2016pagination}. Participants also had more difficulties interacting with the timeline, especiall in regards of selecting in scrolling. This could stem from slightly smaller colliders than the ones in the List, as the elements in the timeline were round and not squared.

The least preferred visualization method was the \textit{Minimap}, which had slightly lower user preference than the \textit{Timeline}. The \textit{Minimap} received significantly higher workload scores, requiring more effort with poorer perceived performance, which led to frustration. Its usability was also significantly lower than that of the other two conditions, with an adjective rating of an \textit{OK} usability \cite{bangor2009sus}. Similarly, in terms of user experience, the \textit{Minimap} was rated significantly worse than the other two conditions, with benchmark comparisons \cite{schrepp2017construction} showing \textit{Below Average} ratings for each scale, except for a \textit{good} rating in novelty. The searching tasks were particularly challenging with the \textit{Minimap}, as only 59.5\% of tasks were completed, and those that were completed took longer. This is in line with our qualitative data, as there werer difficulties users encountered during the searching task in the \textit{Minimap}. The primary issues identified were the challenges in interacting with the 2D POIs on the map, as these were too small for many participants (N=10). This issue was exacerbated by bumpy streets, which made interaction even more difficult. Additionally, the name of the currently hovered POI was displayed on top of the map, leading to the Midas touch problem for some participants (N=9). On a positive note, participants generally appreciated the function and design of the map, and some (N=5) indicated they would have rated it higher if not for the aforementioned issues. Additionally, three participants mentioned that they would prefer a combination of the \textit{List} and \textit{Minimap}, as the \textit{Minimap} was generally liked as a tool for quickly getting an overview, while the \textit{List} was preferred for searching specific items.

To address \textbf{RQ1}, our findings indicate that the \textit{List} visualization technique is the most effective for helping passengers identify missed and upcoming POIs. Therefore, we recommend utilizing a vertical List for POI discovery, as it consistently outperformed other techniques.

\subsection{Interacting via Eye-Gaze and Pinch}
In this Section, we answer \textbf{RQ2$_{main}$:} \textit{Is eye-gaze and pinch a feasible interaction method for interacting with both world-fixed POIs and car-fixed UIs?} mainly through findings from our qualitative interviews. The interaction technique via eye-gaze and pinch proved overall problematic, with the majority of the participants' problems and negative feedback stemming from the interaction technique. This problem was seen in all conditions and, as such was indepentent of the visualization technique. The only feature that exacerbated the issue were the small selectable icons in the \textit{Minimap}, as discussed in Section \ref{section:discussion_passed_upcoming}. We hypothesize, that the main issues stem from using a pinch gesture as selection confirmation and for scrolling, even though we used state-of-the-art technology with the Leap Motion 2\footref{foot:Leapmotion2}. Our pre-study and the related work \cite{Schramm2023Assessing} did not show these issues to the same degre, as they used a hardware-button for confirmation. As such, we suggest to further examine these interaction techniqes, with a preference for using some kind of hardware assistance for selection confirmation and scrolling. E.g. a smartphone as input device could prove useful. When asked for potential additional features for the system, some participants also suggested a multimodal approach for input. For example using voice input when searching for a POI via name, or setting a filter for specific categories via voice.

For \textbf{RQ2}, we conclude that eye-gaze and pinch is not a viable interaction method for engaging with both world-fixed POIs and car-fixed UIs. Instead, we recommend adopting a multimodal approach for in-car AR interfaces, combining head-gaze or eye-gaze with voice and/or hardware-based inputs.

\subsection{Exploring POIs using AR}
In this section, we answer \textbf{RQ3$_{main}$:} \textit{Would users accept an VST-based in-car AR system to explore POIs?} by utilizing the quantitative data presented in Section \ref{section:results} and the findings from our qualitative interviews. Our system received overall positive feedback with, depending on the condition, relatively low workload, good usability, above average to excellent user experience, and low motion sickness scores. Only the minimap condition showed results that are below average across metrics. The system received a moderate score for Essentiality, often categorizing it as a nice system to have, but not essential in its current state. Especially the hardware limitations were often pointed out, as the Varjo XR-3 is rather warm and heavy. Participants could imagine using the system only with small and lightweight ar-glasses or with HUDs in the windshield. As for use-cases, nine participants mentioned wanting to use it during a city trip or road trip in unknown environments for exploration. Six participants can imagine using such a system daily, e.g. to look for a gas station nearby or to display information about ratings or opening hours for nearby locations.

To address \textbf{RQ3}, we conclude that users find VST-based in-car AR systems acceptable for exploring POIs. However, for hardware, we recommend prioritizing compact and lightweight AR glasses or windshield-based AR displays to enhance comfort.

\subsection{Integration into an existing Framework}
Kim and Jung \cite{kim2019automotive} proposed a cognitive framework for passengers in autonomous vehicles using traditional displays. They identified key cognitive needs for passengers, including driving route, current time, estimated duration, parking information, and location suggestions. Their UI framework incorporates four key design components: (A) time tracking, (B) driving purpose, (C) tour guidance, and (D) parking options. Our system utilizes AR to address components (B) and (C) of this framework. For (B), it displays recommended and favorite places in the surrounding area, aligning with the driving purpose. For (C), it provides informational windows with images and detailed descriptions of specific locations, fulfilling the tour guide component. The general acceptance of our AR system among participants further demonstrates its potential to meet the cognitive needs outlined in their framework, suggesting that AR can effectively enhance passenger experiences.


\subsection{Limitations}
In-car AR systems are currently severly limited by available hardware. Current HMDs, such as the Varjo XR-3\footref{foot:Varjo} used in our studies offer many needed functionalities such as eye tracking and hand tracking, but are unwieldy or uncomfortable in return \cite{Goedicke2022xroom}. As such, passengers don't want to wear these all the time. Similarly other, smaller AR-glasses are more comfortable but offer less features and smaller FOVs.

The main problems users faced in our system stemmed from technical issues. Even though we used the Leap Motion 2\footref{foot:Leapmotion2} mounted on the HMD, users had troubles in performing the pinch and drag gestures. Here, factors such as sunlight, handedness, and experience with HMDs can play a role. As the Varjo XR-3 is quite large and heavy, the influence of vehicle movements was also significant. This makes accurate interaction hard, especially when interacting with small UI elements, such as the ones on the minimap.

Additionally, our survey exclusively targeted employees from an automotive software company. While this demographic aligns closely with potential users of such a system and may possess relevant expertise, it could also introduce bias into the survey results.
\section{Conclusion and Future Work}
\label{section:Conclusion}
In summary, this paper demonstrated the potential of in-car Augmented Reality (AR) systems to enhance the passenger experience by providing interaction with the environment through world-fixed Points of Interest (POIs). We show that passengers and drivers have a need for interacting with the environment while on the go and want to seek information about locations in their environment. We also found that both passengers and drivers experience difficulties in recalling names of missed POIs. To find missed POIs, the name, a category, and a description are preferable information. We also show passengers and drivers desire to create lists of interesting locations they meet. Both groups show clear preferences to save POIs manually. Regarding in-car AR interaction, we demonstrate that eye-gaze with a hardware button is a feasible interaction technique for interacting with world-fixed POIs. Conversely, our research also highlights the limitations of the use of eye-gaze combined with pinch gestures in a moving vehicle, which proved problematic for many users. As such, we recommend using interaction techniques that are less prone to movements for in-car AR such as handheld or car-fixed devices. Additionally, we investigated three visualizations for exploring passed and upcoming POIs while traveling, with a clear preference among users for list-based visualizations due to their familiarity and ease of use. The timeline and minimap visualizations, while innovative, require further refinement to address usability challenges, particularly in dynamic environments. The overall system received scores in the mid-range, partly caused by current hardware limitations. Users would prefer lighter hardware or no head-worn devices at all. In general, the integration of AR in vehicles holds great promise for transforming travel time into a more engaging, informative, and productive experience. However, achieving this potential will require continued exploration and refinement of both the hardware technology and the user interfaces that support it.

Future work should explore alternative interaction methods, possibly incorporating multimodal approaches such as voice commands or hardware-assisted inputs to enhance usability and user satisfaction. Additionally, the placement of world-fixed POIs can be improved, e.g. by utilizing computer vision approaches to anchor POIs to buildings. Systems to mitigate vehicle motion for interaction also seem promising. We also want to explore the impact of AR-POIs for various driving scenarios, such as commuting and travel.

\begin{acks}
We thank Stephan Leenders, Oscar Ariza, Axel Hildebrand and Sarah Gökeler for their support and contributions. We also sincerely thank the participants of our studies for their time and valuable input.
\end{acks}

\bibliographystyle{ACM-Reference-Format}
\bibliography{bibliography}

\clearpage

\appendix
\section{Appendix}
\subsection{Data - Survey on current passenger behaviour regarding POIs}

\begin{center}
    \begin{minipage}{\textwidth}

        \centering
        \captionof{table}{Usage of various navigation methods by our survey participants.}
        \begin{tabular}{r|cc|cc|cc|cc|cc}
            \toprule
            & \multicolumn{2}{c|}{AppleCar/AndroidAuto} & \multicolumn{2}{c|}{InVehicleSystems} & \multicolumn{2}{c|}{SmartphoneApps} & \multicolumn{2}{c|}{Compass} & \multicolumn{2}{c}{PaperMaps} \\
            & N & \% & N & \% & N & \% & N & \% & N & \% \\
            \midrule
            Never & 39 & 36.4\% & 13 & 12.1\% & 2 & 1.9\% & 101 & 94.4\% & 79 & 73.8\% \\
            Rarely & 19 & 17.8\% & 19 & 17.8\% & 15 & 14.0\% & 2 & 1.9\% & 26 & 24.3\% \\
            Sometimes & 18 & 16.8\% & 22 & 20.6\% & 33 & 30.8\% & 2 & 1.9\% & 2 & 1.9\% \\
            Often & 22 & 20.6\% & 34 & 31.8\% & 32 & 29.9\% & 0 & 0\% & 0 & 0\% \\
            Always & 9 & 8.4\% & 19 & 17.8\% & 25 & 23.4\% & 2 & 1.9\% & 0 & 0\% \\
            \bottomrule
        \end{tabular}

        \vspace{\baselineskip}
        
        \captionof{table}{The types of information drivers and passengers need to find a missed point of interest. Multiple choice was possible.}
        \begin{tabular}{r|cc|cc}
            \toprule
            \textbf{Type of information} & \multicolumn{2}{c|}{Drivers} & \multicolumn{2}{c}{Passengers} \\
            & N & \% & N & \% \\
            \midrule
            Name & 71 & 89\% & 65 & 90\% \\
            Perspective Picture & 39 & 49\% & 38 & 53\% \\
            Web Picture & 46 & 58\% & 44 & 61\% \\
            Description Text & 45 & 56\% & 49 & 68\% \\
            Category & 45 & 56\% & 49 & 68\% \\
            \bottomrule
        \end{tabular}

        \vspace{\baselineskip}

        \captionof{table}{Percentages on how and when passengers and drivers look for a missed point of interest.}
        \begin{tabular}{r|c|c}
            \toprule
            \textbf{Action}             & \textbf{Drivers}   & \textbf{Passengers} \\
            \midrule
            Nothing                     & 5\%                 & 4\% \\
            System saves Automatically  & 0\%                 & 3\% \\
            Stop the Car                & 3\%                 & 0\% \\
            Search later on Smartphone  & 46\%                & 8\% \\
            Search immediately on Smartphone & 11\%           & 49\% \\
            Search later on Navigation system & 24\%          & 5\% \\
            Search immediately on Navigation system & 11\%    & 31\% \\
            \bottomrule
        \end{tabular}

    \end{minipage}
\end{center}

\clearpage
\subsection{Data - Pre-Study on Eye-Gaze Interaction}

\begin{center}
    \begin{minipage}{\textwidth}
      
    \centering
    \captionof{table}{Descriptive Statistics for the pre-study Raw NASA Task Load Index.}
    \begin{tabular}{l|c|c|c|c|c|c|c}
        \toprule
        & \textbf{Mental} & \textbf{Physical} & \textbf{Temporal} & \textbf{Performance} & \textbf{Effort} & \textbf{Frustration} & \textbf{Score} \\
        \midrule
        Mean & 15.0 & 19.5 & 40.0 & 33.5 & 28.0 & 13.5 & 24.8 \\
        Median & 12.5 & 15.0 & 42.5 & 25.0 & 32.5 & 15.0 & 25.0 \\
        Standard Deviation & 10.5 & 19.9 & 23.1 & 21.4 & 18.1 & 11.6 & 9.47 \\
        Shapiro-Wilk W & 0.942 & 0.782 & 0.961 & 0.848 & 0.898 & 0.882 & 0.959 \\
        Shapiro-Wilk p & 0.573 & 0.009 & 0.794 & 0.055 & 0.206 & 0.139 & 0.769 \\
        \bottomrule
    \end{tabular}  

    \vspace{\baselineskip}

    \captionof{table}{Descriptive Statistics for the pre-study System Usability Scale.}
    \begin{tabular}{l|c|c|c|c|c}
        \toprule
        \textbf{Question} & \textbf{Mean} & \textbf{Median} & \textbf{Std. Deviation} & \textbf{Shapiro-Wilk W} & \textbf{Shapiro-Wilk p} \\
        \midrule
        Frequent Use & 4.00 & 4.00 & 0.943 & 0.841 & 0.045 \\
        Unnecessary Complex & 1.40 & 1.00 & 0.516 & 0.640 & < .001 \\
        Easy to Use & 4.60 & 5.00 & 0.516 & 0.640 & < .001 \\
        Support of Technical & 1.50 & 1.00 & 0.972 & 0.603 & < .001 \\
        Well Integrated & 4.40 & 4.00 & 0.516 & 0.640 & < .001 \\
        Inconsistency & 1.80 & 2.00 & 0.789 & 0.820 & 0.025 \\
        Learn Quickly & 4.60 & 5.00 & 0.699 & 0.650 & < .001 \\
        Cumbersome & 1.60 & 1.00 & 0.966 & 0.678 & < .001 \\
        Confident & 4.20 & 4.00 & 0.632 & 0.794 & 0.012 \\
        Learn a Lot Before & 1.10 & 1.00 & 0.316 & 0.366 & < .001 \\
        Score & 86.0 & 87.5 & 8.01 & 0.925 & 0.398 \\
        \bottomrule
    \end{tabular}

    \end{minipage}
\end{center}

\clearpage
\subsection{Data - Study on Visualizing Passed and Upcoming POIs}

\begin{center}
    \begin{minipage}{\textwidth}

        \centering
        \captionof{table}{Descriptive Statistics for the visualization-study Raw NASA Task Load Index, grouped by condition.}
        \begin{tabular}{llccccccc}
            \toprule
            \textbf{Statistic} & \textbf{Condition} & \textbf{Mental} & \textbf{Physical} & \textbf{Temporal} & \textbf{Performance} & \textbf{Effort} & \textbf{Frustration} & \textbf{Score} \\
            \midrule
            \multirow{3}{*}{Mean} & List & 28.6 & 25.5 & 20.7 & 14.8 & 29.5 & 19.5 & 23.1 \\
            & Timeline & 29.8 & 27.1 & 23.3 & 20.0 & 27.4 & 28.6 & 26.0 \\
            & Minimap & 46.7 & 36.4 & 42.4 & 42.6 & 56.9 & 50.2 & 45.9 \\
            \midrule
            \multirow{3}{*}{Median} & List & 25 & 20 & 15 & 5 & 25 & 15 & 25.0 \\
            & Timeline & 25 & 20 & 15 & 5 & 20 & 20 & 22.5 \\
            & Minimap & 40 & 30 & 45 & 40 & 60 & 55 & 49.2 \\
            \midrule
            \multirow{3}{1.5cm}{Std. Deviation} & List & 19.8 & 21.3 & 17.0 & 19.5 & 25.3 & 18.5 & 14.6 \\
            & Timeline & 22.8 & 21.3 & 19.6 & 29.0 & 26.6 & 26.0 & 19.3 \\
            & Minimap & 21.5 & 24.7 & 22.3 & 22.8 & 23.5 & 22.6 & 15.8 \\
            \midrule
            \multirow{3}{1.5cm}{Shapiro-Wilk W} & List & 0.917 & 0.879 & 0.909 & 0.767 & 0.835 & 0.851 & 0.941 \\
            & Timeline & 0.868 & 0.904 & 0.911 & 0.683 & 0.821 & 0.874 & 0.927 \\
            & Minimap & 0.947 & 0.886 & 0.920 & 0.938 & 0.954 & 0.946 & 0.927 \\
            \midrule
            \multirow{3}{1.5cm}{Shapiro-Wilk p} & List & 0.076 & 0.014 & 0.052 & < .001 & 0.002 & 0.004 & 0.229 \\
            & Timeline & 0.009 & 0.043 & 0.058 & < .001 & 0.001 & 0.011 & 0.118 \\
            & Minimap & 0.297 & 0.019 & 0.088 & 0.201 & 0.399 & 0.290 & 0.118 \\
            \bottomrule
        \end{tabular}

        \vspace{\baselineskip}

        \captionof{table}{Descriptive Statistics for the visualization-study System Usability Scale scores, grouped by condition.}
        \begin{tabular}{lccccc}
            \toprule
            \textbf{Condition} & \textbf{Mean} & \textbf{Median} & \textbf{Std. Deviation} & \textbf{Shapiro-Wilk W} & \textbf{Shapiro-Wilk p} \\
            \midrule
            List & 78.5 & 77.5 & 10.2 & 0.885 & 0.018 \\
            Minimap & 61.1 & 60.0 & 14.3 & 0.962 & 0.563 \\
            Timeline & 71.5 & 75.0 & 15.9 & 0.921 & 0.091 \\
            \bottomrule
        \end{tabular}

        \vspace{\baselineskip}

        \caption{Descriptive Statistics for the visualization-study Motion Sickness Questionnaire, grouped by time of completion.}
        \begin{tabular}{lccccc}
            \toprule
            \textbf{Study Order} & \textbf{Mean} & \textbf{Median} & \textbf{Std. Deviation} & \textbf{Shapiro-Wilk W} & \textbf{Shapiro-Wilk p} \\
            \midrule
            Pre-study & 0.381 & 0 & 0.669 & 0.617 & < .001 \\
            After First Condition & 0.857 & 0 & 1.15 & 0.732   & < .001 \\
            After Second Condition & 1.19 & 1 & 1.36 & 0.822   & 0.001  \\
            After Third Condition & 1.00 & 1 & 1.05 & 0.826    & 0.002  \\
            \bottomrule
        \end{tabular}

    \end{minipage}
\end{center}

\begin{table*}[ht]
    \centering
    \caption{Descriptive Statistics for the visualization-study User Experience Questionnaire, grouped by condition.}
    \begin{tabular}{llcccccc}
    \toprule
    \textbf{} & \textbf{Condition} & \textbf{Attractiveness} & \textbf{Perspicuity} & \textbf{Efficiency} & \textbf{Dependability} & \textbf{Stimulation} & \textbf{Novelty} \\
    \midrule
    \multirow{3}{*}{\textbf{M}} & Timeline & 1.26 & 1.42 & 1.10 & 1.12 & 1.11 & 1.37 \\
    & Minimap & 0.952 & 0.964 & 0.619 & 0.821 & 1.00 & 1.57 \\
    & List & 1.78 & 2.05 & 1.79 & 1.68 & 1.54 & 1.31 \\
    \midrule
    \multirow{3}{*}{\textbf{Mdn}} & Timeline & 1.50 & 1.75 & 1.25 & 1.25 & 1.25 & 1.50 \\
    & Minimap & 1.00 & 0.750 & 0.500 & 0.500 & 1.00 & 1.25 \\
    & List & 1.83 & 2.25 & 1.50 & 1.50 & 1.50 & 1.25 \\
    \midrule
    \multirow{3}{*}{\textbf{SD}} & Timeline & 0.921 & 1.03 & 0.937 & 0.883 & 0.986 & 1.11 \\
    & Minimap & 0.972 & 1.01 & 0.993 & 0.946 & 0.939 & 1.13 \\
    & List & 0.642 & 0.692 & 0.755 & 0.717 & 0.704 & 0.798 \\
    \midrule
    \multirow{3}{*}{\textbf{W}} & Timeline & 0.938 & 0.833 & 0.951 & 0.912 & 0.970 & 0.933 \\
    & Minimap & 0.972 & 0.975 & 0.980 & 0.948 & 0.967 & 0.877 \\
    & List & 0.976 & 0.893 & 0.847 & 0.926 & 0.942 & 0.982 \\
    \midrule
    \multirow{3}{*}{\textbf{p}} & Timeline & 0.200 & 0.002 & 0.359 & 0.059 & 0.737 & 0.160 \\
    & Minimap & 0.783 & 0.844 & 0.918 & 0.313 & 0.661 & 0.013 \\
    & List & 0.866 & 0.026 & 0.004 & 0.112 & 0.235 & 0.953 \\
    \bottomrule
    \end{tabular}
\end{table*}

\begin{table*}[ht]
    \centering
    \caption{User ranking values for the visualization-study conditions.}
    \begin{tabular}{l|cc|cc|cc}
    \toprule
    \multirow{2}{*}{\textbf{Condition}}& \multicolumn{2}{c|}{\textbf{Least Favorite}} & \multicolumn{2}{c|}{\textbf{Middle}} & \multicolumn{2}{c}{\textbf{Favorite}} \\
                & N     & \%             & N    & \%                  & N       & \%                 \\
    \midrule
    List        & 2     & 9.5\%          & 4    & 19.0\%              & 15      & 71.4\%              \\
    Minimap     & 11    & 52.4\%         & 7    & 33.3\%              & 3       & 14.3\%             \\
    Timeline    & 8     & 38.1\%         & 10   & 47.6\%              & 3       & 14.3\%             \\
    \bottomrule
    \end{tabular}
\end{table*}

\end{document}